\documentclass[12pt]{article}
\pdfoutput=1
\usepackage[centertags]{amsmath}
\usepackage[square, comma, sort&compress,numbers]{natbib}
\usepackage{array,multirow}
\numberwithin{equation}{section}
\usepackage{amssymb}
\usepackage{graphicx}
\usepackage{color}
\usepackage{relsize}
\usepackage[utf8]{inputenc}

\usepackage{epsfig}

\def\tr{{\rm Tr}}

\def\ep{\epsilon}

\def\Or[#1]{{\text{O}}\left({#1}\right)} 
\def\dotl[#1,#2]{\left\langle #1, #2 \right\rangle}
\def\dotlb[#1,#2]{[ #1, #2 ]}
\def\dotp[#1,#2]{(#1) \cdot (#2)}
\def\aff[#1,#2]{\hat{#1}(#2)}
\def\n4sym{{\cal N}=4 SYM}
\def\>{\rangle}
\def\<{\langle}
\def\weight[#1,#2,#3]{\{(#1),#2,#3\}}
\def\ads[#1]{$\text{AdS}_{#1}$}

\newcommand{\ba}{\begin{eqnarray}}
\newcommand{\ea}{\end{eqnarray}}
\newcommand{\be}{\begin{eqnarray}}
\newcommand{\ee}{\end{eqnarray}}
\newcommand{\bq}{\begin{equation}}
\newcommand{\eq}{\end{equation}}
\newcommand{\benn}{\begin{equation*}}
\newcommand{\eenn}{\end{equation*}}
\newcommand{\bi}{\begin{itemize}}  
\newcommand{\ei}{\end{itemize}}

\newcommand{\CL}{{\cal L}}

\newcommand{\CO}{{\cal O}}

\newcommand{\CV}{{\cal V}}
\newcommand{\bz}{{\bar z}}
\newcommand{\pd}{{\partial}}
\newcommand{\nn}{\nonumber}

\newcommand\oo\infty
\newcommand\s\sigma
\newcommand\de\delta
\newcommand\De\Delta

\newcommand\f\phi
\newcommand\g\gamma
\newcommand\x\times

\usepackage{jheppub}

\makeatletter
\def\@fpheader{\vspace{-.1cm}}
\makeatother

\title{Exact Virasoro Blocks from Wilson Lines \\  
and Background-Independent Operators}

\author[a]{A.\ Liam Fitzpatrick,}
\author[b]{Jared Kaplan,}
\author[b]{Daliang Li,}
\author[b,c]{Junpu Wang}
\affiliation[a]{Department of Physics, Boston University, \\
Commonwealth Avenue, Boston, MA 02215, U.S.A.}
\affiliation[b]{Department of Physics and Astronomy,  Johns Hopkins University, \\
Charles Street, Baltimore, MD 21218, U.S.A.} 
\affiliation[c]{Department of Physics, Yale University, New Haven, CT 06511}

\abstract{  
Aspects of black hole thermodynamics and information loss can be derived as a consequence of Virasoro symmetry.  To bolster the connection between Virasoro conformal blocks and  AdS$_3$ quantum gravity, we study sl$(2)$ Chern-Simons Wilson line networks and revisit the idea that they compute a variety of CFT$_2$ observables, including Virasoro OPE blocks, exactly.  We verify this in the semiclassical large central charge limit and to low orders in a perturbative $1/c$ expansion.  

Wilson lines connecting the boundary to points in the bulk play a natural role in bulk reconstruction.  Because quantum gravity in AdS$_3$ is rigidly fixed by Virasoro symmetry, we argue that sl$(2)$ Wilson lines provide building blocks for background independent bulk reconstruction.  In particular, we show explicitly that they automatically `know' about the uniformizing coordinates appropriate to any background state.  
} 

\arxivnumber{} 

\begin{document}   
  
\maketitle    
\flushbottom

\section{Introduction and Summary}  

In the late 1980s, long before the discovery of the full AdS/CFT correspondence, it was recognized that aspects of $2+1$ dimensional Chern-Simons theory have a $1+1$ dimensional CFT interpretation.  This line of thinking originated with Witten's study of knot theory \cite{Witten:1988hf}, and has had an enormous impact on rational conformal field theory \cite{Elitzur:1989nr, Moore:1989vd}, condensed matter theory \cite{Zhang:1988wy, Moore:1991ks}, quantum computing \cite{Kitaev:1997wr}, and quantum gravity \cite{Witten:1988hc, Witten:1989sx, Verlinde:1989ua}.  The subject's early literature \cite{Witten:1988hf, Verlinde:1989ua} already made the point that conformal blocks emerge from the quantization of Chern-Simons theories.   We would like to make this as precise and explicit as possible for the case of the Virasoro conformal blocks, which are the atomic constituents of CFT$_2$ correlators.   

\begin{figure}[t!]
\begin{center}
\includegraphics[width=0.16\textwidth]{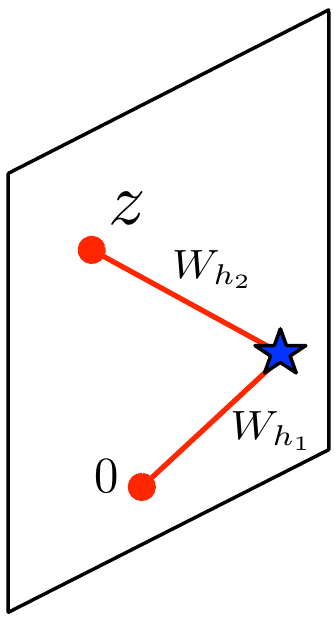} \hspace{8.5mm}
\includegraphics[width=0.28\textwidth]{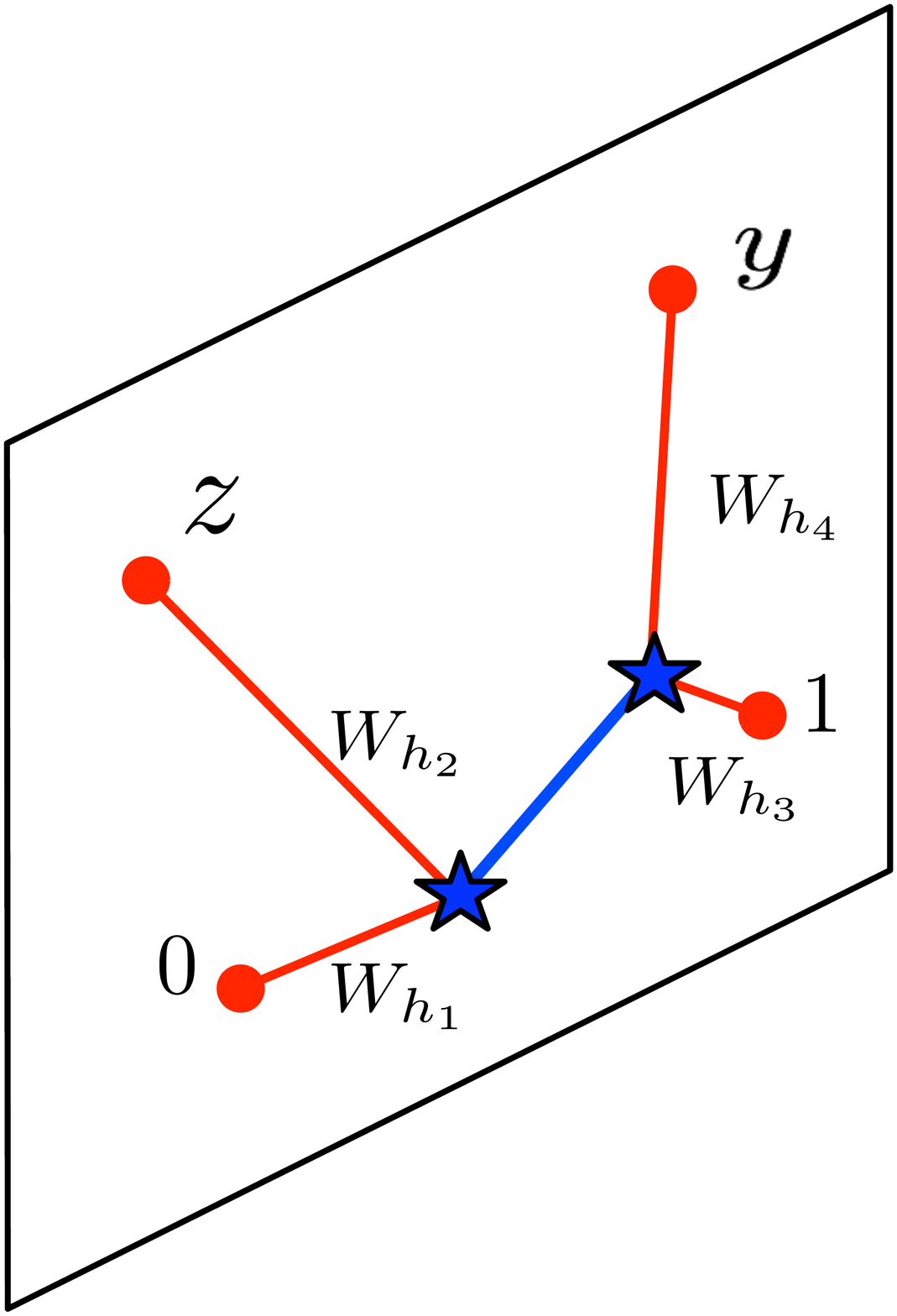} \hspace{4mm}
\includegraphics[width=0.28\textwidth]{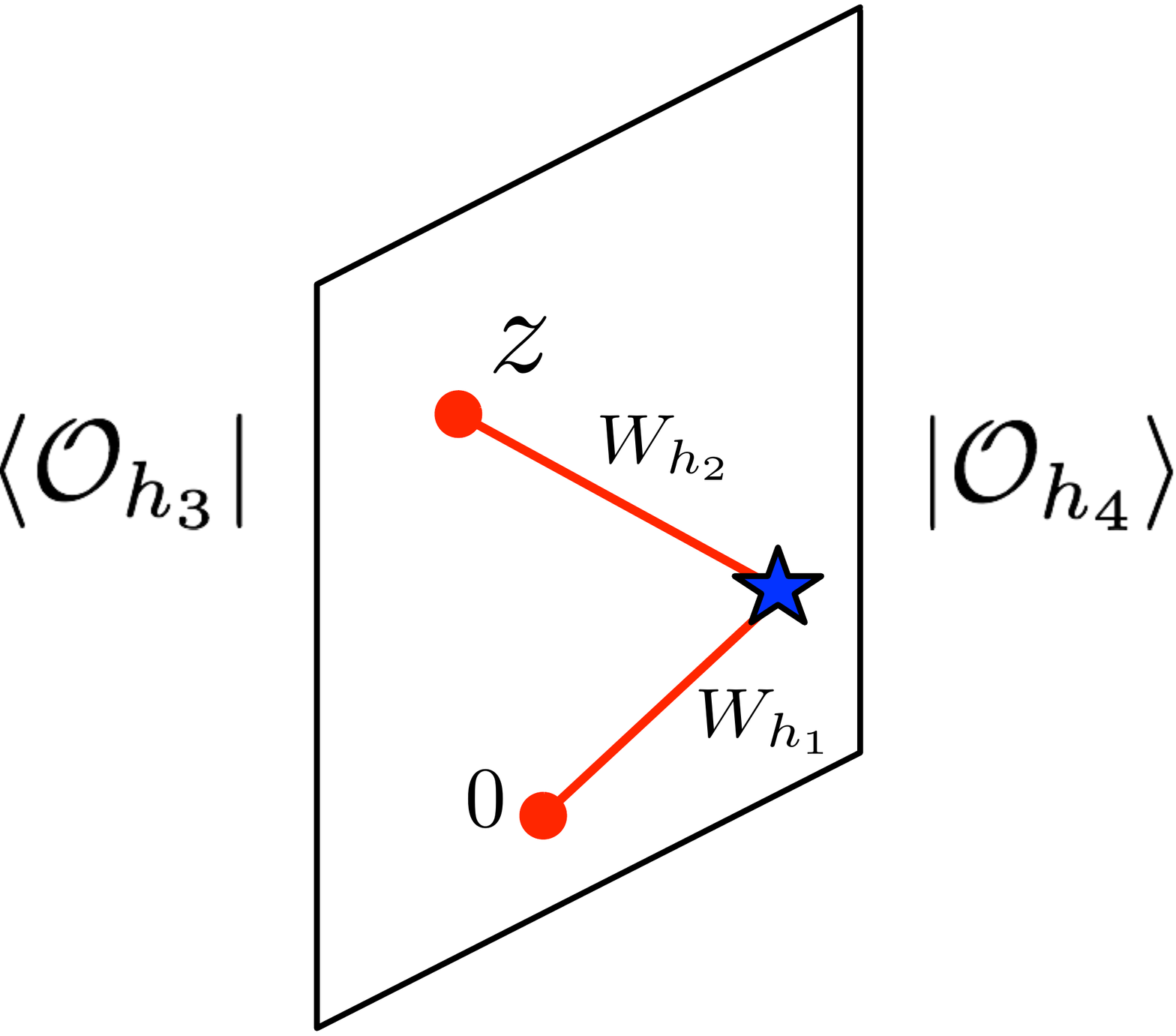}
\caption{{\it Left:} A sketch of Wilson lines computing a Virasoro OPE block.  {\it Middle:} By putting together two such OPE blocks, one obtains a Wilson line network computing a Virasoro conformal block. The blue line indicates the non-trivial vacuum expectation value of the product of the OPE blocks.  {\it Right:} The Virasoro conformal blocks can also be computed by putting the OPE blocks in the appropriate background bra and ket states. }
\label{fig:NetworkforIntro}
\end{center} 
\end{figure}

We will study Wilson lines transforming in infinite-dimensional representations of sl$(2)$ \cite{Verlinde:1989ua} and propagating in a three-dimensional half-space that can be identified with AdS$_3$.\footnote{For more early works involving Chern-Simons Wilson lines applied to 3d gravity, see e.g. \cite{carlip1989exact,de1990spin,vaz1994wilson}. For more recent works using Chern-Simons Wilson lines to compute entanglement entropy, see e.g. \cite{Ammon:2013hba,deBoer:2013vca, Hijano:2014sqa, deBoer:2014sna, Hegde:2015dqh, Chen:2016uvu, Chen:2016kyz}}  We will see that combinations of such Wilson lines can explicitly construct the full contribution to the OPE  from an irreducible Virasoro representation.\footnote{The name ``OPE Blocks''  has been proposed \cite{Czech:2016xec} for irrep contributions to the OPE.}  Putting together multiple such contributions produces certain Wilson line networks, such as that pictured in figure \ref{fig:NetworkforIntro}, that compute the Virasoro conformal blocks.  Roughly speaking, this follows because this network satisfies the Virasoro Ward identity for a local correlator \cite{Verlinde:1989ua}, and because the individual Wilson lines propagate states \cite{Witten:1988hf, Elitzur:1989nr} in irreducible representations of Virasoro.  These ideas have appeared in the literature before, but recent explicit calculations \cite{KrausBlocks, Hijano:2015qja, Besken:2016ooo, Alkalaev:2015wia, Alkalaev:2015lca, Hulik:2016ifr} did not venture beyond the semiclassical limit, and early work on the subject was rather formal and implicit.

Our interest in the `bulk' or Chern-Simons description of Virasoro blocks has been motivated by two recent developments:  progress in relating Virasoro blocks to AdS$_3$ black hole thermodynamics \cite{Fitzpatrick:2014vua, Fitzpatrick:2015zha} and information loss \cite{Fitzpatrick:2016ive, Fitzpatrick:2016mjq, Anous:2016kss}, and by renewed interest in bulk reconstruction, especially beyond black hole horizons
 \cite{Almheiri:2012rt, Papadodimas:2012aq, Papadodimas:2013jku, Almheiri:2014lwa}.  

Black hole thermodynamics, the Cardy formula, eigenstate thermalization, and various notions of information loss all arise as a consequence of the behavior of the Virasoro blocks in the semiclassical large central charge or $c \to \infty$ limit  \cite{Fitzpatrick:2014vua, Fitzpatrick:2015zha, Fitzpatrick:2016ive, Fitzpatrick:2016mjq, Anous:2016kss}.  Furthermore, one can go beyond this limit and explicitly calculate non-perturbative effects\footnote{From the viewpoint of eigenstate thermalization, these effects transcend the thermodynamic limit.}  of the parametric form `$e^{-c}$' within the structure of the blocks.  These effects alter or resolve some of the information loss problems. The existence of a `bulk' or `gravitational' formalism that exactly computes the Virasoro blocks (and not only their semiclassical limit \cite{KrausBlocks, Hijano:2015qja, Besken:2016ooo, Alkalaev:2015wia, Alkalaev:2015lca}) suggests that the gravitational path integral in AdS$_3$ may have a precise meaning.  This would be a remarkable statement about AdS$_3$ quantum gravity. 
A sharp definition for the gravitational path integral should be a boon for those who seek to reconstruct (and thereby define!) the bulk.  In section \ref{sec:VersionsofReconstruction} we will break the bulk reconstruction question into several sub-problems, and then we will explain why special features of AdS$_3$/CFT$_2$ provide a unique line of attack.  The essential point is that Virasoro symmetry completely determines many aspects of AdS$_3$ quantum gravity, making it plausible to hope for concrete non-perturbative predictions concerning bulk reconstruction.\footnote{Some closely related ideas have been suggested recently \cite{daCunha:2016crm, Lewkowycz:2016ukf, Guica:2016pid}.  While some of the motivations are similar, our proposal appears to differ in detail from \cite{Lewkowycz:2016ukf}, but our methods are closely connected with the very recent work of Guica \cite{Guica:2016pid}.}

\subsection{Kinematics and Dynamics in Bulk Reconstruction}
\label{sec:VersionsofReconstruction}

We would like to understand how to reconstruct AdS physics from CFT data and dynamics.  Let us try to break this hard and very general problem  into a few conceptually separate pieces.  The simplest version of the problem is to study reconstruction only in the limit $G_N=0$ in a pure AdS background.  From this starting point, we can move toward the complexity of the general problem by keeping $G_N=0$ but allowing a fixed asymptotically AdS background (e.g. AdS Schwarzschild).  Alternatively, we can keep a pure AdS background but allow a fluctuating quantum gravitational geometry on top of it, with $G_N \neq 0 $.  These two directions have some overlap with each other, since one can build classical geometries as coherent states of fluctuations around the vacuum.  Ultimately, one would like to consider a fully quantum geometry with an arbitrary expectation value and an exact description of the fluctuations. 

We should also decide whether our goal for bulk  reconstruction will be to reproduce only  kinematical structure, i.e. if it will be based entirely on the symmetries of the theory, or whether we want to account for dynamics.  We will begin by explaining these perspectives.  Then we will discuss reconstruction in the AdS$_3$/CFT$_2$ context, where the reconstruction of the bulk kinematic structure is remarkably rich and includes much of the non-linear behavior of quantum gravity. 

To begin with, consider an interacting non-gravitational QFT in pure AdS$_{d+1}$.  
The standard way to associate a naive bulk operator $\phi^{(0)}(X)$ with every CFT primary operator $\CO(x)$ is to use an integral transform often referred to as the HKLL kernel \cite{Hamilton:2005ju,Witten:1998qj, Banks:1998dd}.  These $\phi^{(0)}$ are natural kinematical constituents for interacting bulk operators, but if the bulk theory has any interactions, the procedure must be modified to obtain the correct $n$-pt correlators in the bulk. 

Including the effect of such bulk interactions requires that we grapple with the more difficult issue of dynamics. For perturbative interactions, one can go beyond the trivial kinematical approximation of the $\phi^{(0)}$ by working order-by-order in perturbation theory and imposing causality and unitarity constraints on the bulk and bulk-boundary correlators.  This process has seen significant study \cite{Hamilton:2006az, Kabat:2011rz, Kabat:2012av, Kabat:2016zzr}, but it is not entirely clear if and when it provides a unique, well-defined algorithm.  In any case, it would be nice to know whether a well-defined procedure exists for the case when the (non-gravitational) bulk QFT has no perturbative expansion parameters.\footnote{Along these lines, an  interesting and potentially non-perturbative proposal based on the bulk-boundary OPE was recently suggested \cite{Paulos:2016fap}.    Their idea is to define a Euclidean quantization in the bulk by using the AdS dilatation operator to expand and contract hemispheres surrounding a point on the boundary of AdS.    In cases where the AdS QFT happens to be conformal, this would make AdS/CFT identical to the bulk/boundary CFT relationship (see \cite{Liendo:2012hy} for a contemporary discussion).  Nevertheless, when the AdS QFT is not conformal, it may not have an OPE, so it is unclear what consistency conditions we should impose to well-define the bulk reconstruction.  But this formulation  may provide a precise and tractable starting point, as has been emphasized to us by M. Paulos. }

Ultimately, we want to go beyond bulk reconstruction in pure AdS$_{d+1}$.  Once we deform the bulk geometry, we cannot use symmetries to relate $\phi(X)$ to $\CO(x)$, even if our bulk fields are non-interacting.  In other words, even the simplest version of bulk reconstruction will no longer be purely kinematical.  We can still proceed by working backwards to construct modified HKLL kernels from an analysis of bulk QFTs, but the procedure is not manifestly well-defined.    More to the point, this process is {\it incomplete} from the boundary CFT point of view -- the gravitational background should not be input into the formalism by hand.  Instead, the geometry should be derived as an output that appears automatically when we compute correlators of reconstructed bulk operators within an excited, high-energy CFT state.  This reconstruction problem also has important conceptual differences \cite{Papadodimas:2013jku, Almheiri:2014lwa} from the non-gravitational case.

In AdS$_3$/CFT$_2$ it is possible to resolve many of these issues using the power of the Virasoro algebra.  
Much of the quantitative behavior of quantum gravity in AdS$_3$ arises as a consequence of Virasoro symmetry \cite{Fitzpatrick:2015zha, Fitzpatrick:2016ive, Fitzpatrick:2016mjq}.   Non-trivial gravitational backgrounds, including those of black holes, can be observed to emerge automatically from Virasoro `kinematics'.  We would like to put these ideas to work, taking advantage of the fact that AdS$_3$ quantum gravity is particularly rigid and well-defined.

As a practical matter, one would like to define operators $\Phi^{(0)}$ that have a natural interpretation as the kinematical constituents of a local bulk field \emph{in any background state} \cite{Guica:2016pid}.  The $\Phi^{(0)}$ will have a geometric interpretation whenever bulk geometry is a meaningful concept.  So the $\Phi^{(0)}$ can be interpreted as vast generalizations of the global conformal $\phi^{(0)}$, which only have a nice interpretation in the vacuum.   Furthermore, the $\Phi^{(0)}$ should have an exact definition (up to gauge transformations), so their behavior can be meaningfully analyzed in setups where non-perturbative quantum gravitational effects become important.  Our formalism makes it possible to analyze these operators in detail \cite{Czech:2016xec, Guica:2016pid} and to study their implications for the black hole information paradox, insofar as this is possible without accounting for non-gravitational bulk dynamics.

\subsection{Summary and Outline} 

We study correlators of sl$(2)$ Wilson lines in an infinite dimensional representation, where the generators take the form of 
equation (\ref{eq:RepresentationofL}).  In this basis, a Wilson line from $Z_i$ to $Z_f$ can be written as the `matrix'
\be
W_h(Z_f; Z_i) = \int dx \, |  x \> P \left\{ e^{\int_{Z_i}^{Z_f} d Z^\mu A^a_\mu(z) L^a_x} \right\} \< x | .
\ee
When $Z_i$ and $Z_f$ are attached to the boundary at $y=0$, as pictured in figure \ref{fig:WilsonLineAndNetwork}, the Wilson line will be a gauge invariant observable transforming under the Virasoro algebra as a correlator of local operators located at the endpoints.  
 Wilson lines can also connect via gauge invariant vertices to form networks, and examples like that of figure \ref{fig:NetworkforIntro} compute Virasoro conformal blocks to all orders in $1/c$.  We will see how to explicitly compute Wilson line networks with endpoints on the boundary by moving the lines themselves to the boundary, so that their primary matrix elements take the simple form
\be
\< h | W_h(z_f; z_i) | h\> = \left. P \left\{ e^{\int_{z_i}^{z_f} d z \left[ \partial_x + \frac{12}{c} T(z) \left( \frac{1}{2} x^2 \partial_{x} + h x \right) \right] } \right\} \frac{1}{x^{2h}} \right|_{x = 0} .
\ee
The stress tensor $T(z)$ appearing in this equation is an operator, and not just a classical field.   Thus Virasoro blocks can be computed in $1/c$ perturbation theory in terms of integrals over multi-stress tensor correlators,\footnote{In appendix \ref{app:ReviewCSandHolography} we review how stress tensor correlators can be derived from Chern-Simons theory.} as we demonstrate in section \ref{sec:ComputingUsingWilsonLines}.    

In the presence of a uniformly continuous background $\<T(z)\>$, 
one can locally find a uniformizing $w(z)$ coordinate system such that $\< T(w) \> =0$ 
once we transform the CFT to the non-trivial background $ds^2 = dw d \bar w$.  Ignoring global issues and singularities, the uniformizing coordinates can be extended into AdS$_3$ \cite{Asplund:2014coa,Banados:1998gg} to produce a metric 
\be \label{AAdS3 metric}
&\to& \frac{dy^2}{y^2} +  \left(\frac{1}{y^2} + \frac{y^2}{4} L(z) \bar L(\bar z) \right) dz d \bar z + \frac{L(z)}{2} dz^2 + \frac{\bar L(\bar z) }{2}d \bar z^2,  
\ee
where $L(z) =  -\frac{12}{c} T(z)$.
This metric automatically satisfies the vacuum Einstein's equations in the presence of the energy-momentum sources $\CO_i(z_i)$.

As we discuss in section \ref{sec:PathIntegralFormalismandReconstruction}, the Wilson lines can be re-written in a path-integral form that manifests a striking connection with uniformizing coordinates. In a general background, the  Wilson line can be expressed as
\be
\< h | W(z_f; z_i)|h\> = \left( e^{\int_{z_i}^{z_f} dz \frac{12  T(z)}{c} x_T(z)}\frac{1}{x_T(z_i)^{2}}\right)^h ,
\ee
where the path-integral constrains $x_T(z_i)$  to be the solution to an equation of motion involving the stress tensor,
\be
-x_T'(z) &=& 1 + \frac{6 T(z)}{c} x_T^2(z), \qquad x_T(z_f)=0.
\ee
  In particular, the equation of motion is satisfied by
\be 
\frac{1}{x_T(z)} &=& \frac{w''(z)}{2w'(z)}- \frac{w'(z)}{w(z)- w(z_f)},
\ee 
where $w(z)$ obey the uniformizing coordinate condition of equation (\ref{eq:UniformizingCoords}) at an operator level, ie as a functional of the operator $T(z)$.  Then the primary matrix element of a Wilson line becomes 
 \be
\< h | W_h(z_f; z_i) | h\> &=& \left( \frac{w'(z_f) w'(z_i) }{(w(z_f) - w(z_i))^2} \right)^h.
\ee
Many similar results follow for Wilson line networks and OPE blocks, demonstrating that the Wilson lines compute higher point correlators and conformal blocks correctly.  In particular, the non-vacuum Virasoro OPE blocks can be constructed by ``dressing'' the global OPE blocks with Wilson lines.  That is, integral expressions for the global OPE blocks can be written simply in terms of kernels $f(z_1,z_2, z_3)$:
\be
\CO_1(z_1) \CO_2(z_2) \supset \int_{z_1}^{z_2} dz_3 f(z_1, z_2, z_3) \CO_3(z_3) ,
\ee
where these kernels can be derived using the ``shadow field'' formalism \cite{Ferrara:1971vh,Ferrara:1973vz,Ferrara:1972xe,Ferrara:1972uq,Ferrara:1972ay,SimmonsDuffin:2012uy}.  To obtain the Virasoro blocks, one simply uses a modified ``quantum'' kernel $F$ instead of $f$, where $F$ depends non-linearly on the stress tensor through Wilson lines connecting the operators $\CO_1, \CO_2$ to the operator $\CO_3$:
\be
F(z_1, z_2, z_3) &=&\int dx_1 dx_2 W_{h_1} (z_1; 0 ; z_3, x_1) W_{h_2}(z_2; 0; z_3, x_2) f(x_1, x_2, 0).
\ee
Precise definitions and computations of this object are given in the body of the paper, as well as explicit checks in both the $1/c$ expansion and in the semi-classical limit in an arbitrary background.

The outline of this paper is as follows.  In section \ref{sec:WilsonLines}, we define the Wilson lines and discuss how to use them to construct  Virasoro OPE blocks that satisfy the Virasoro Ward identities.  In section \ref{sec:ComputingUsingWilsonLines}, we study a perturbative large $c$ expansion and show that our formalism reproduces known results; importantly, we verify terms that represent quantum $\frac{1}{c}$ corrections beyond the semi-classical limit.  In section \ref{sec:PathIntegralFormalismandReconstruction}, we discuss a representation of the Wilson line using a  path integral whose fundamental degrees of freedom reside in an internal space associaed with infinite-dimensional representations of the conformal algebra. Using this representation we show how the uniformizing coordinates for a general background are automatically computed by the Wilson lines and reproduce the semi-classical limit of the vacuum and non-vacuum blocks.   Section \ref{sec:Discussion} provides a discussion, while in appendix \ref{app:ReviewCSandHolography} we review  aspects of sl$(2)$ Chern-Simons theory relevant to AdS/CFT,  in appendix \ref{app:PerturbationTheoryandShadows} we collect some technical details, and in appendix \ref{app:Regulation} we discuss the regulation of divergences.

\section{Chern-Simons Wilson Lines and CFT$_2$}
\label{sec:WilsonLines}

Our main goal is to show how a prescription for Virasoro conformal blocks in CFT$_2$ arises from a Chern-Simons formulation of AdS$_3$ gravity.  The connection between Virasoro blocks and sl$(2)$ Chern-Simons Wilson lines was first articulated by Verlinde in a prescient 1989 paper \cite{Verlinde:1989ua}, following up on related results on WZW models \cite{Witten:1988hf, Elitzur:1989nr} and the Chern-Simons description of AdS$_3$ gravity \cite{Witten:1988hc}.  We will modernize Verlinde's prescription and adapt it to the usual AdS/CFT setup where CFT operators can be taken to `live on the boundary' of an AdS spacetime.\footnote{Older literature \cite{Witten:1988hf, Verlinde:1989ua, Witten:1988hc, Elitzur:1989nr} outlines a prescription, but provides few explicit computations.  In more recent work the semiclassical Virasoro blocks \cite{Hijano:2015qja, Alkalaev:2015wia, Alkalaev:2015lca, Besken:2016ooo} have been successfully obtained from AdS gravity and Chern-Simons theory, but it was unclear how these methods could be extended to compute the Virasoro blocks to all orders in $1/c$.  Our goal here is an exact and explicit prescription.  }

The atomic objects we construct along the way will in fact be more versatile: we will see how the Chern-Simons formulation naturally leads to compact expressions for the partial contributions from Virasoro irreps to the OPE of two primary operators, i.e. the ``Virasoro OPE Blocks'':
\be
\left[ \CO_1(z) \CO_2(0) \right]_{\CO_p \textrm{ irrep}} = \frac{C_{12p}}{z^{h_1+h_2-h_p}} \CO_p(0) + \textrm{descendants},
\ee
where all Virasoro descendants of the primary operator $\CO_p$ are included in the sum.  
One can use these OPE blocks to compute the contribution of the irreducible representation corresponding to $\CO_p$ in a general correlator or background state.

AdS/CFT provides our primary motivation, so let us briefly recall the Chern-Simons description of AdS$_3$ gravity; we provide a more complete review in appendix \ref{app:ReviewCSandHolography}.  We can decompose the bulk metric in terms of a pair of sl$(2)$ gauge fields as
\be
g_{\mu \nu} = \tr \left[ (A - \bar A )_\mu (A - \bar A )_\nu \right] 
\ee
where $A$ and $\bar A$ are interpreted as $2 \times 2$ matrices in sl$(2)$.  The gravitational action is the difference $I_{CS}[A] - I_{CS}[\bar A]$, where the Chern-Simons action is
\be
\label{eq:CSAction}
I_{\rm CS}[A]
&=& \frac{k}{4\pi} \int_{y \geq 0} d z d \bar z \,d y\,\tilde{\epsilon}^{\mu \nu \lambda}
\tr \left(A_\mu \partial_\nu A_\lambda+\frac{2 }{3} A_\mu A_\nu A_\lambda \right)
\ee
and the level $k = \frac{R_{AdS}}{4 G_N} = \frac{c}{6}$ in terms of the bulk parameters or the CFT$_2$ central charge.  To obtain the  Virasoro asymptotic symmetry algebra (rather than an sl$(2)$ Kac-Moody algebra) as the asymptotic symmetry, we must impose the boundary condition
\be
\left. A_z \right|_{y = 0} = L^1 + \frac{12}{c} T(z) L^{-1} .
\label{eq:WilsonLineNearBoundary}
\ee
for $A$, and an equivalent anti-holomorphic condition for $\bar A$.  We will be focusing on the holomorphic sector governed by $A$ throughout this paper.

\subsection{Defining sl$(2)$ Wilson Lines}
\label{sec:DefiningWilsonLines}

\begin{figure}[t!]
\begin{center}
\includegraphics[width=0.85\textwidth]{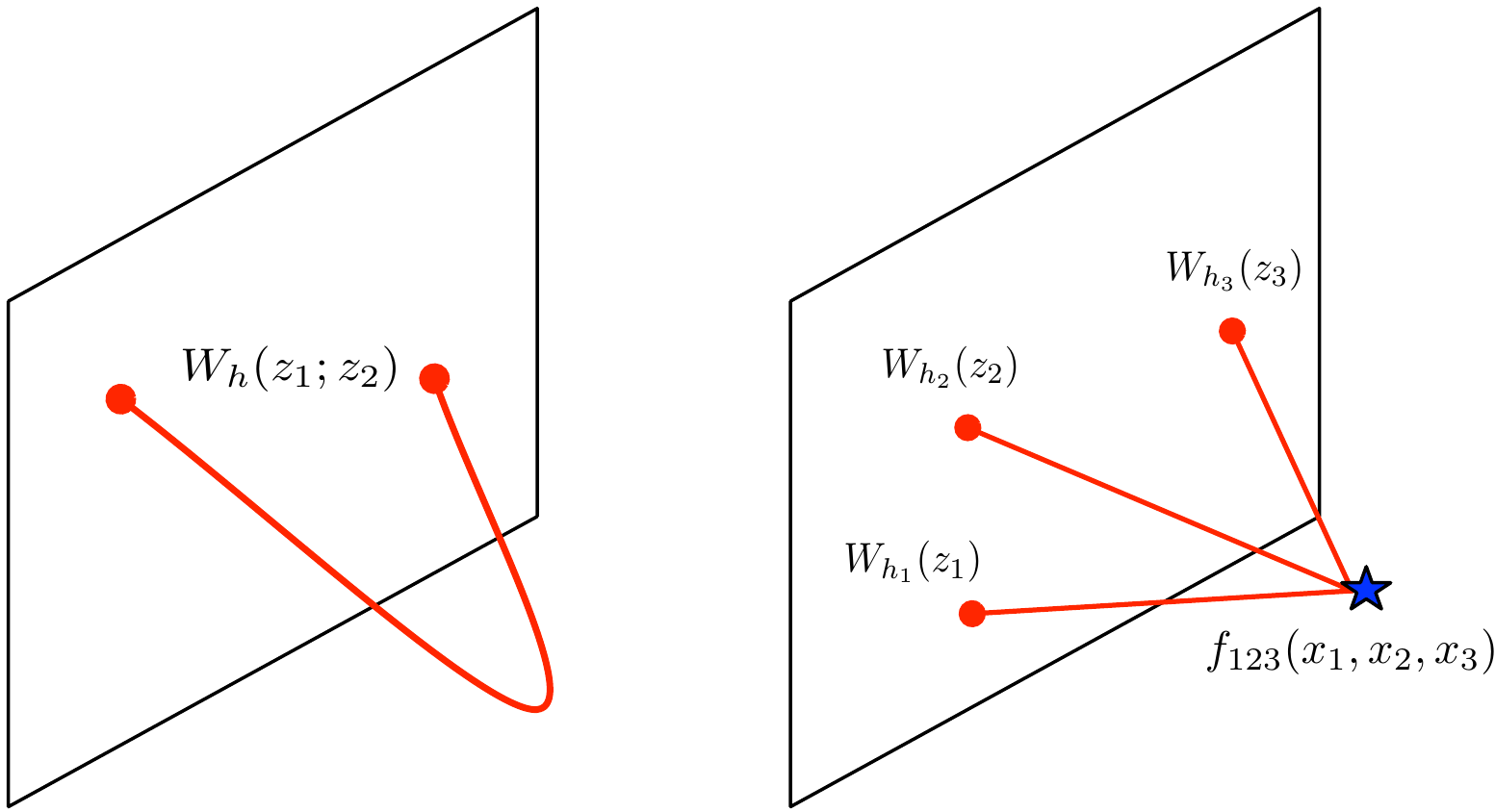}
\caption{This figure shows a single Wilson line $W_h$ ending at two points on the boundary, and a network of three Wilson lines emanating from three boundary points and meeting at a bulk vertex.  The vertex must be invariant under the bulk sl$(2)$ gauge group, and so as a function of the internal sl$(2)$ variables $x_i$ it must take the functional form of a conformally invariant 3-point correlator.   }
\label{fig:WilsonLineAndNetwork}
\end{center}
\end{figure}

To compute OPEs and Virasoro blocks, we will study networks of Wilson lines that end on the boundary, as pictured in figure \ref{fig:WilsonLineAndNetwork}.  We take the endpoints to be `charges'  in infinite dimensional representations of sl$(2)$, chosen to transform like local CFT$_2$ operators under the conformal group.    It will be convenient to choose an explicit basis for the states in these infinite dimensional representations. A possible choice is the discrete basis
\be
| h \>, \ \  L_{-1} | h \>, \ \  L_{-1}^2  | h \>, \ \  L_{-1}^3  | h \>, \cdots  
\ee
where $| h \>$ is a primary state of sl$(2)$, which simply means that it has $L_0$ eigenvalue $h$ and is annihilated by the lowering operator $L_1$.  

We will find another basis  more convenient.  We can take advantage of the fact \cite{Verlinde:1989ua, Ammon:2013hba} that the infinite dimensional  representations of sl$(2)$ can be encoded on the space of holomorphic functions of one auxiliary variable $x$.  When acting on holomorphic functions $\psi(x)$, the sl$(2)$ generators then take the form
\be
L^{1} &\cong & L_{-1}= \partial_{x}
\nn \\
L^0 &\cong & L_0 = x \partial_{x} + h 
\nn \\
L^{-1} &\cong & L_{1} = \frac{1}{2} x^2 \partial_{x} + h x
\label{eq:RepresentationofL}
\ee
where the $L^a$ depend on the holomorphic dimension or weight $h$ that labels the representation.  The path-ordered Wilson lines ending on the boundary of AdS$_3$ 
\be
W_h(z_i; z_f) = P \left\{ e^{\int_{z_i}^{z_f} d z A^a_z(z) L^a} \right\}
\ee
will be infinite-dimensional `matrices' in this sl$(2)$ representation.  That is to say, the  states $|h\>, \dots$ that these matrices act on parameterize the irreps associated with the insertion of the primary operators $\CO(z)$ at the ends of the Wilson lines.  An element of the matrix $W_h$ is simply  (the irrep contribution to) the correlator of two descendant operators:
\be
\left[ W_h(z_i; z_f) \right]_{ij} = \left\< \Big(\partial_z^i \CO_2(z_f)\Big) P \left\{ e^{\int_{z_i}^{z_f} d z A^a_z(z) L^a} \right\} \Big(\partial_z^j \CO_1(z_i)\Big) \right\> .
\ee
 This is almost the standard basis $|h\>, L_{-1} |h\>, \dots$, but as should be clear from the above discussion, the corresponding primary and descendant operators act at $z_i$ and $z_f$ rather than $z=0$, so the basis has been translated from the origin.  

To define the basis in terms of the auxiliary parameter $x$, we must specify the wavefunctions $\psi(x) \equiv \<x | h\>$ for the lowest weight state $|h\>$.  The $L^a$ act on functions $\psi(x)$ in exactly the same way that holomorphic global conformal generators act on correlators.  
Demanding that $L_1$ annihilates the lowest weight state uniquely determines the sl$(2)$ `wavefunction' 
for the primary state vector $| h \>$ written in the $x$-basis to be
\be
\< x | h \> \equiv \frac{1}{x^{2h}} .
\ee
The formal sl$(2)$ space operator $\<   x |$ acts as a projector onto the $x$-basis.  One can easily compute the wavefunctions of specific descendant states by inserting $L_{-1}^k$ into the simple correlators above.  We emphasize that these `correlators' do not reside in the physical space of the CFT$_2$, but only within the auxiliary internal sl$(2)$ space.  

When we study physical CFT$_2$, we usually conjugate by inversions, but for the internal sl$(2)$ we will simply define $\< h | x \> = \delta(x)$.   This construction is reminiscent of the `shadow' representation (see \cite{SimmonsDuffin:2012uy} for a recent discussion).   At a formal level, one writes a shadow operator $\tilde \CO_h(x)$ as
\be
\tilde \CO_h(x) = \int dy \, (x-y)^{2h-1} \, \CO_h(y),
\ee
so that $\< \tilde \CO_h(x) \CO_h(y) \> = \delta(x-y)$. We provide a very explicit review of how shadow fields can be used to project onto global conformal irreps in  appendix \ref{app:ShadowDetails}.   

Now, the auxiliary coordinate $x$ acts like a typical coordinate in quantum mechanics.  Up to a normalization,
\be
{\bf 1}_h = \int dx \, |  x \> \< x | 
\ee
is a projector onto the sl$(2)$ representation with dimension $h$.  A trivial Wilson line will be equal to this `matrix' ${\bf 1}_h$.  The general Wilson line can be written as
\be
\label{eq:DefinitionWilsonLine}
W_h(z_f; z_i) = \int dx \, |  x \> P \left\{ e^{\int_{z_i}^{z_f} d z A^a_z(z) L^a_x} \right\} \< x | .
\ee
Note that this has the desirable composition property 
\be
W_h(c,b) W_h(b,a) = W_h(c,a)  .
\ee
In fact, the Wilson line is an evolution operator in the `time' coordinate $z$.  The path-ordering is just `time' ordering, and the Hamiltonian for evolution in $z$ is just the integrand of the exponential:
\be
H(z) &=& i A_z^a(z) L_x^a .
\ee
Promoting $x$ to an operator $X$ on the auxiliary space, its conjugate momentum is $P = -i \partial_x$, and we can write the Hamiltonian as
\be
H(z) &=&  \left( A_z^{-1}(z) \left(i h X - \frac{1}{2} X^2 P \right) + A_z^0(z) (i h - X P) - A_z^{1}(z) P \right).
\label{eq:WilsonLineHamiltonian}
\ee
States evolve in $z$ according to this Hamiltonian:
\be
\frac{1}{i} \frac{\partial}{\partial z} |\psi; z\> = H(z) |\psi;z\>,
\ee
and the Wilson line  evolves states in $z$
\be
\< \psi_f ; z_i | W_h(z_f; z_i) | \psi_i; z_i \> &=& \< \psi_f; z_f | \psi_i; z_i \> ,
\ee
and so is a kind of propagator.  We can also think of the Wilson lines as functions of $x_f$ and $x_i$ variables via
\be
\label{eq:xDefinitionWilsonLine}
W_h(z_f, x_f; z_i, x_i) &\equiv& \< x_f; z_i | W_h(z_f; z_i) | x_i; z_i\> =  \< x_f; z_f | x_i; z_i \>,
\ee
which is just the usual definition of an evolution operator in quantum mechanics, written in the $x$-basis.

We will often be interested in Wilson lines sandwiched between primary sl$(2)$ states
\be
\< h | W_h(z_i; z_f) | h \> &=& \int dx \, \<h |  x \> P \left\{ e^{\int_{z_i}^{z_f} d z A^a_z(z) L^a_x} \right\} \<  x | h \>
\nn \\
&=& \int dx \, \delta(x)  P \left\{ e^{\int_{z_i}^{z_f} d z A^a_z(z) L^a_x} \right\} \frac{1}{ x^{2h} } .
\ee
In the definition of equation (\ref{eq:xDefinitionWilsonLine}), this arises from integrating against the wavefunctions $\psi_f(x_f)= \delta(x_f)$ and $\psi_i(x_i) = \frac{1}{x_i^{2h}}$ for the bra and ket states.  
As a first example, let us see what happens if we evaluate this Wilson line in the limit that $c \to \infty$.  We chose boundary conditions for the $2+1$ dimensional Chern-Simons field so that 
\be
\left. A_z \right|_{y \to 0} = L^1 + \frac{12}{c} T(z) L^{-1} .
\ee
Wilson lines that lie entirely in the boundary surface at $y=0$, evaluated in large central charge limit $c = \infty$ (with other paramters fixed) simply correspond to evolution with
\be
H &=& - P.
\ee
Consequently, at $c=\infty$, evolution in $z$ is trivial:
\be
\< h; z_f | x \> \stackrel{c=\infty}{=} \<h  | e^{ -i P (z_f-z_i)}| x  \> = \< h  |  x-( z_f- z_i) \> = \delta(x-(z_f-z_i)).
\label{eq:InfCEvol}
\ee
Overlapping with the initial state $|h \> = |h; z_i\> $ produces
\be
\< h; z_f | h; z_i \> = \int dx \< h; z_f| x \> \< x | h\> \stackrel{c=\infty}{=} \frac{1}{(z_f-z_i)^{2h}}.
\ee
Equivalently, we can see this directly in terms of the $c=\infty$ path-ordered Wilson line:
\be
\< h | W_h^{c=\infty}(z_i; z_f) | h \>  &=& \int dx \, \delta(x)  e^{\int_{z_i}^{z_f} d z \, \partial_x} \frac{1}{ x^{2h} }
=\frac{1}{(z_f-z_i)^{2h}} .
\ee
We see explicitly that the inner product of wavefunctions has been traded for a primary operator 2-pt function that depends on the physical spacetime coordinates $z_i$.  

More generally, the Virasoro Ward identity imposes constraints that transform the internal coordinate $x$ into a physical coordinate.  In effect, the Wilson lines promote internal sl$(2)$ transformations into physical Virasoro transformations.  This is equivalent to the `dressing' of charged fields by Wilson lines in other gauge theories.  We will discuss the Virasoro Ward identity in section \ref{sec:VirasoroWardIdentity}, with a full derivations in appendix \ref{app:ReviewCSandHolography}.

\subsection{Virasoro OPE Blocks from Wilson Line Networks}
\label{sec:OPEsfromWilsonLines}

We saw in section \ref{sec:DefiningWilsonLines} how to define an sl$(2)$ Wilson line as an infinite dimensional matrix labeled by internal space coordinates $x_i$.  We can form more general operators by contracting  the $x$-space labels of several Wilson lines with sl$(2)$-invariants.  

Consider the setup pictured on the right in figure \ref{fig:WilsonLineAndNetwork}, where three Wilson lines emanating from $z_1, z_2, z_3$ meet at a point in the bulk. Schematically, near the bulk vertex $Z$ the Wilson lines take the form
\be
e^{ \int^Z A^a L^a_{x_1} +  \int^Z A^a L^a_{x_2} +  \int^Z A^a L^a_{x_3}}  f_{123}(x_1, x_2, x_3)
\ee
Bulk gauge invariance under the infinitesimal transformation $A_\mu^a \to A_\mu^a + \partial_\mu \phi^a$ implies 
\be
\phi^a(Z) \left( L_{x_1}^a +  L_{x_2}^a + L_{x_3}^a \right) f_{123}(x_1, x_2, x_3)  = 0
\ee
for any $\phi^a(Z)$, which requires $f_{123}$ to take the form of a conformally invariant 3-pt correlator in $x$-space 
\be
f_{123} \propto \frac{1}{x_{12}^{h_1 + h_2 - h_3 } x_{23}^{h_2 + h_3 - h_1} x_{31}^{h_3 + h_1 - h_2} } .
\label{eq:shadow3point}
\ee
We can use these vertices to construct gauge-invariant Wilson line networks.  As an example, formal arguments from section \ref{sec:VirasoroWardIdentity} suggest that the network pictured in the center of figure \ref{fig:NetworkforIntro} should compute a Virasoro conformal block.

However instead of focusing on Virasoro blocks for correlators, let us construct a Virasoro OPE block using these Wilson lines.  To begin with, note that in CFT$_2$ a global conformal OPE block \cite{Czech:2016xec} can be written
\be
\CO_1(z_1) \CO_2(z_2) \supset N \int_{z_1}^{z_2} dz_3 f_{12,\tilde 3}( z_1, z_2, z_3) \CO_3(z_3) .
\label{eq:GenOPEBlock1}
\ee
where the shadow dimension $\tilde h_3 = 1 - h_3$ has replaced $h_3$ in $f_{12 \tilde 3}$ and $N$ is a normalization factor.\footnote{Taking $N = \frac{\Gamma(2h_3)}{\Gamma(h_3 + h_{12})\Gamma(h_3 - h_{12})}$,  reproduces a standard convention for the normalization of conformal blocks.}  We verify this formula explicitly using the shadow formalism \cite{Ferrara:1971vh,Ferrara:1973vz,Ferrara:1972xe,Ferrara:1972uq,Ferrara:1972ay,SimmonsDuffin:2012uy} in appendix \ref{app:ShadowDetails}.

We will use Wilson lines to construct a more general operator $F_{12,3}$ computing a Virasoro OPE block 
\be
\CO_1(z_1) \CO_2(z_2) \supset N \int dz_3 F_{12,3}( z_1, z_2;z_3) \CO_3(z_3) .
\ee
The idea is to promote global conformal symmetry to the full Virasoro symmetry by `dressing' the correlator with Wilson lines.  Roughly speaking, for this purpose we can add the structure on the right of figure \ref{fig:WilsonLineAndNetwork}, simplifying a bit by moving the bulk vertex to the boundary point $z_3$.  We can also take the Wilson lines to lie entirely on the boundary and take advantage of the condition \ref{eq:WilsonLineNearBoundary} to write the Chern-Simons field in terms of the stress tensor.
At finite $c$, the operator $F_{12,3}(z_1, z_2; z_3)$ becomes a ``quantum'' kernel that depends non-linearly on the stress tensor.  In terms of Wilson lines, it is
\be
\label{eq:OPEfromWilsonLines}
F_{12,3} = \int d x_1 dx_2  W_{h_1}(z_1; 0; z_3, x_1) W_{h_2}(z_2; 0; z_3, x_2) f_{12 \tilde 3}(x_1, x_2, 0) ,
\ee
where we have set some indices $x \to 0$ to identify the Wilson line endpoints as primary operators.
Using the $c=\infty$ expression (\ref{eq:InfCEvol}) for the Wilson line evolution, it is trivial to evaluate $F_{12,3}$ explicitly in this limit and observe that $F_{12,3}$ reduces to $f_{12 \tilde{3}}$:
\be
F_{12,3}^{c=\infty} &=& \int d x_1 dx_2  \, \delta(x_1 - (z_1-z_3) ) \delta(x_2- (z_2-z_3))  f_{12 \tilde 3}(x_1 , x_2 ,0) 
\nn \\
&=&   f_{12 \tilde 3}( z_{13}, z_{23}, 0) = f_{1 2 \tilde 3}(z_1, z_2, z_3) .
\ee
As we will argue in the remainder of this work, at finite $c$ our Wilson line formalism from equation (\ref{eq:OPEfromWilsonLines})  computes the full Virasoro OPE block.  At a formal level, this should follow because our OPE block obeys the Virasoro ward identity and propagates the correct states.  But we will also compute the OPE block explicitly, at both the semiclassical level and at the quantum level in $1/c$ perturbation theory.

\begin{figure}[t!]
\begin{center}
\includegraphics[width=0.35\textwidth]{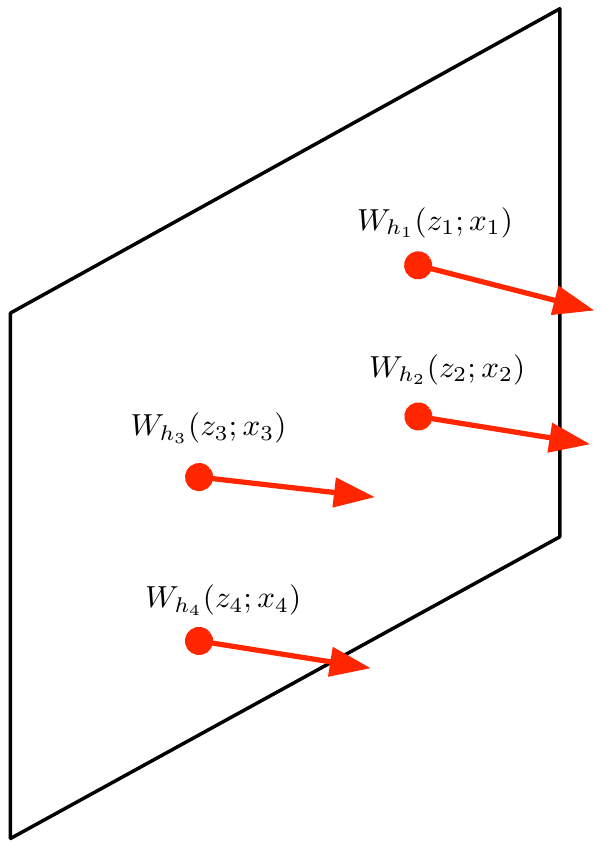}
\caption{This figure shows some Wilson lines anchored near the boundary at $y=0$ at various points $z_i$ and pointing into the bulk.  The Wilson lines are labelled by an $x_i$ variable transforming in the infinite dimensional representation of sl$(2)$ with primary dimension $h_i$. From the near-boundary behavior we deduce that this obeys the Virasoro Ward identity for a correlator of primary operators with dimensions $h_i$ located at $z_i$.}
\label{fig:SomeWilsonLinesNearTheBoundary}
\end{center}
\end{figure}

\subsection{The Virasoro Ward Identity and Chern-Simons Hilbert Spaces}
\label{sec:VirasoroWardIdentity}

\begin{figure}[t!]
\begin{center}
\includegraphics[width=0.55\textwidth]{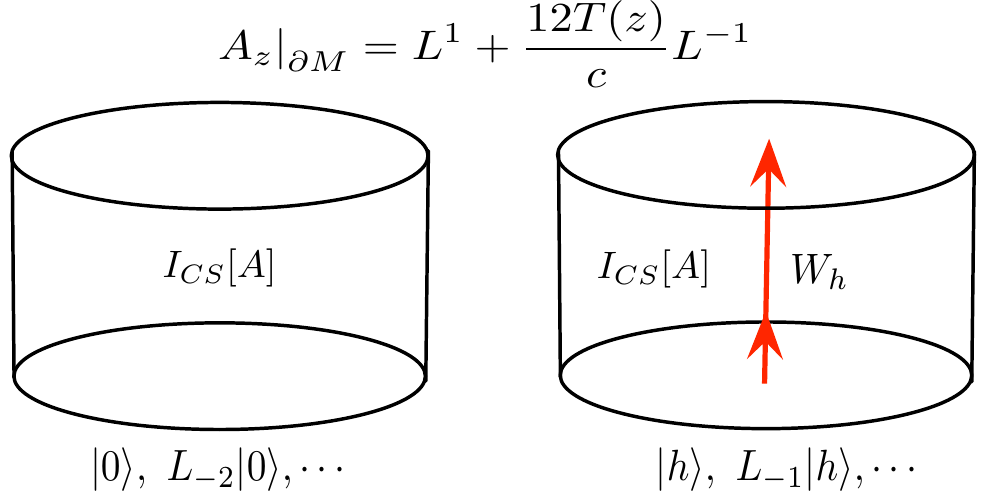}
\caption{The Hilbert space associated with an empty cylinder consists of the vacuum and its Virasoro descendants.  When we include a Wilson line $W_h$, the space of states includes  all Virasoro descendants of a dimension $h$  primary state.  }
\label{fig:WilsonLineHilbertSpace}
\end{center}
\end{figure}

Two motivations for our construction are Ward identities and the Hilbert space of sl$(2)$ Chern-Simons theory in the presence of Wilson lines.  The Wilson line correlators discussed in the previous two sections obey a version of the  Virasoro Ward identity \cite{Verlinde:1989ua}.  Absent subtleties from regularization, this implies that Wilson line correlators compute linear combinations of Virasoro conformal blocks   \cite{Witten:1988hf, Elitzur:1989nr, Verlinde:1989ua}.  Quantizing Chern-Simons theory on a time-slice punctured by a Wilson line produces a Hilbert space consisting of an irreducible representation of the Virasoro algebra associated with that Wilson line, as suggested in figure \ref{fig:WilsonLineHilbertSpace}.  This gives a simple interpretation for the intermediate states in Wilson line networks.

Let us first review the Virasoro Ward identity for CFT correlators.  It is convenient to state the identity in terms of a generating functional
\be
\Psi[\mu; z_i] &=& \left\< \CO_1(z_1) \dots \CO_N(z_N) e^{i \int d^2 z \mu(z, \bar{z})) T(z)} \right\>
\ee
for the correlator of some Virasoro primaries $\CO_i$ and any number of stress tensor insertions.
The identity takes the form
\be
&& \left( \bar{\partial} - \mu(z) \partial - 2 (\partial \mu(z) ) \right) \left( \frac{\delta}{\delta \mu(z)} \Psi[\mu; z_i] \right) + \frac{c}{12}  \frac{\partial^3 \mu(z)}{2 \pi i} \Psi[\mu; z_i]
\nn\\
&& =  \sum_i \left( h_i \partial \delta^2(z-z_i) + \delta^2(z-z_i) \partial_{z_i} \right) \Psi[\mu; z_i] .
\label{eq:StatementVerlindeTypeWardIdentity} 
\ee
Note that $\frac{\delta}{\delta \mu(z)}$ brings down a factor of the stress tensor $T(z)$, and so the delta function terms on the second line arise from contact terms between the stress tensor and the other operators in the correlator.   In section \ref{sec:ComputingUsingWilsonLines} we will need to regulate certain stress-tensor correlators, but as long as we preserve these contact terms, the Ward identity will be preserved.  

Any function obeying the Virasoro Ward identity will transform correctly under the full two-dimensional conformal group, and so it can be viewed as a candidate CFT$_2$ correlator.  In particular, such functions will have a decomposition in Virasoro conformal blocks.  We can determine which specific blocks appear by examining which states appear in the OPE.

Wilson line correlators are governed by a version of the Virasoro Ward identity, as Verlinde \cite{Verlinde:1989ua} first showed.  Gauge invariant  Wilson line correlators can only include Wilson lines with endpoints on the boundary at $y=0$, as pictured in figure \ref{fig:SomeWilsonLinesNearTheBoundary}.  The Wilson lines emanating from the boundary can connect up in a variety of gauge-invariant ways, but we do not need to specify this information in order to derive the Virasoro Ward identity.  Thus in place of primary operators, we include a Wilson line $W_{h_i}(z_i, x_i) | h_i \>$.  Dropping the $|h_i\>$ for notational simplicity, we have
\be
\Psi[\mu; z_i, x_i] &=& \left\< W_{h_1}(z_1, x_1) \dots W_{h_n}(z_n, x_n) e^{i \int d^2 z \mu(z, \bar{z})) T(z)} \right\> .
\ee
We review two very different  derivations of the  Ward identity for these Wilson line correlators in appendix \ref{app:VirasoroWardfromCS}.  The first is based on holographic renormalization \cite{deHaro:2000vlm}, while the latter follows Verlinde's \cite{Verlinde:1989ua} use of the gauge constraints.  As an important consequence of these Ward identities, we learn that
\be
\left( \partial_{x_i} - \partial_{z_i} \right) \Psi[\mu; z_i, x_i] = 0
\ee
for each pair of $x_i$ and $z_i$.  Thus the Ward identity requires us to identify the internal sl$(2)$ coordinate space parameterized by the $x_i$ with the physical spacetime coordinates $z_i$ of the Wilson line endpoints.  

These arguments demonstrate that our Wilson line correlators must compute some linear combination of Virasoro conformal blocks.  To see that they compute individual blocks associated with specific states, we need to understand the space of states associated with a propagating Wilson line.  In fact, this question was also addressed long ago \cite{Witten:1988hf, Witten:1988hc, Witten:1989sx, Verlinde:1989ua, Elitzur:1989nr, Moore:1989vd}.  When we quantize sl$(2)$ Chern-Simons theory in a cylinder pierced by a Wilson line as suggested in figure \ref{fig:WilsonLineHilbertSpace}, the Hilbert space corresponds to the Virasoro primary state associated with the Wilson line's representation and all of its Virasoro descendants.  For completeness, we review the quantization of Chern-Simons theory in the vacuum in appendix \ref{app:ReviewCSandHolography}.

This provides an interpretation for Wilson line networks \cite{Moore:1989vd}.  If we `slice' the network in such a way that our time slice includes only a Wilson line in a representation $h$, we can interpret the network as computing a sum over intermediate states in the irrep labeled by $h$.  If our slice includes two Wilson lines, the Hilbert space includes a tensor product of the two representations.  It is also possible to interpret the `monodromy method' or `accessory parameter method' for computing semiclassical Virasoro blocks in Chern-Simons theory by studying Wilson loop linking \cite{deBoer:2014sna}.

\section{Wilson Line Correlators in $1/c$ Perturbation Theory}
\label{sec:ComputingUsingWilsonLines}

In this section we will use our formalism to compute Virasoro blocks in $\frac{1}{c}$ perturbation theory.  For both the vacuum and general Virasoro blocks, we work to one order  beyond the semiclassical limit in the large central charge expansion, and verify that our results match with previous computations.  This provides evidence that our Wilson line construction is an exact definition of the Virasoro OPE blocks.

\subsection{Vacuum Block at Order $\frac{1}{c^2}$}

In this section we will explain how correlators of Wilson lines restricted to the $y=0$ boundary surface can be computed straightforwardly in terms of the $n$-point correlation functions of the CFT$_2$ stress tensor.  
The latter can be calculated using the Chern-Simons description, or using well-known recursion relations for $\< T(z_n) \cdots T(z_1) \>$ correlators; we review the computation of stress tensor correlators from Chern-Simons theory in appendix \ref{app:ReviewCSandHolography}. Using these results it is straightforward to evaluate Wilson line correlators in $1/c$ perturbation theory.  In this section we will compute the Virasoro vacuum block to order $\frac{1}{c^2}$, explicitly demonstrating that our formalism works beyond the semiclassical limit.

\begin{figure}[t!]
\begin{center}
\includegraphics[width=0.85\textwidth]{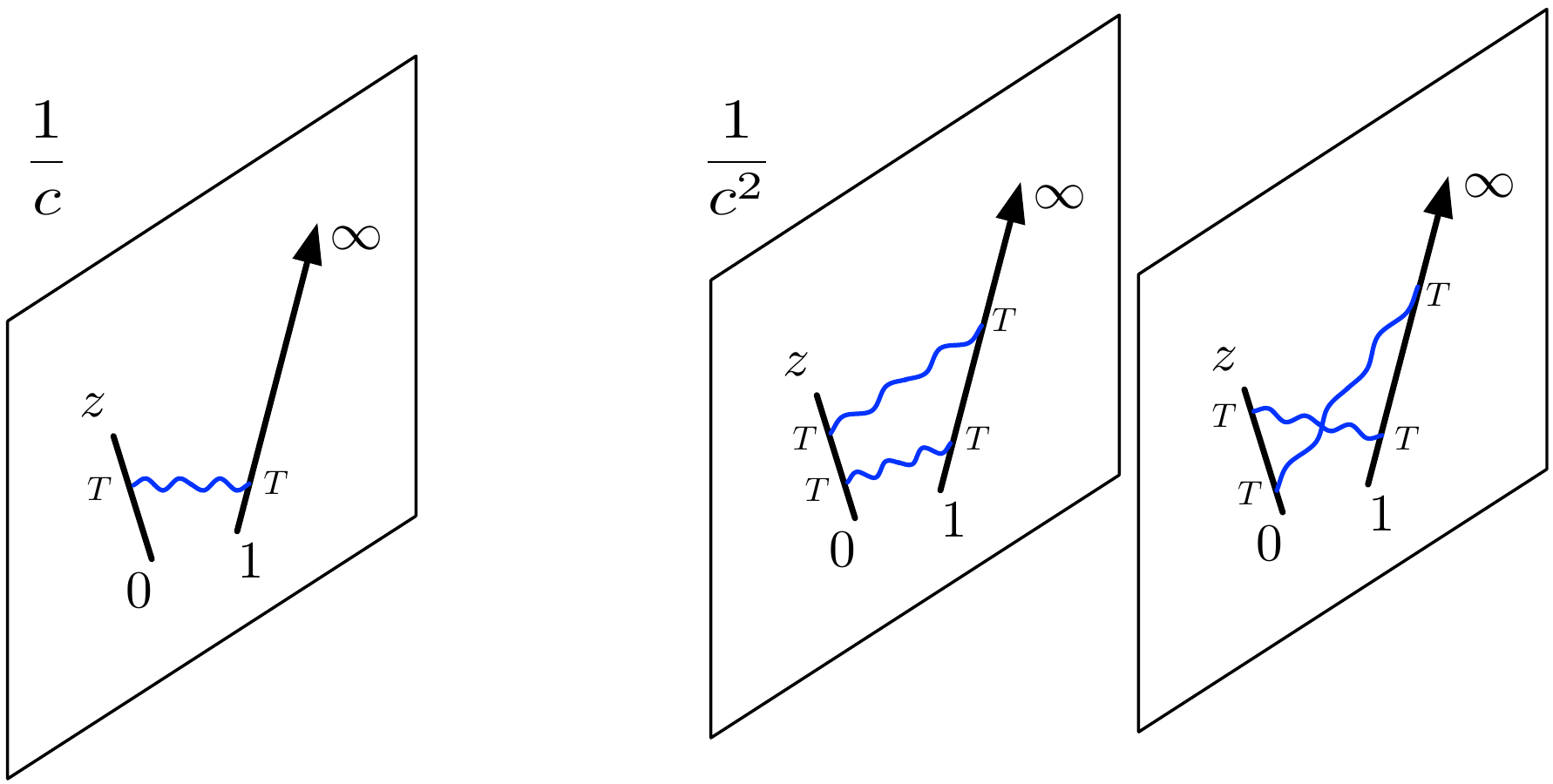}
\caption{This figure indicates the contributions to the Virasoro vacuum block at order $\frac{1}{c}$ and $\frac{1}{c^2}$.  The Wilson lines appear in black, while the wavy blue lines indicate contractions of stress energy tensors $\< TT \> = \frac{c}{2 z_{ij}^4}$.  ``Connected'' stress tensor correlators begin to contribute at order $\frac{1}{c^3}$. 
 }
\label{fig:PerturbativeVacuumBlock}
\end{center}
\end{figure}

As a starting point, we simply note that as a consequence of the boundary condition in equation (\ref{eq:WilsonLineNearBoundary}), we can write Wilson lines that propagate along the boundary as
\be
W_h(z_f; z_i) = \int dx \, | x \> P \left\{ e^{\int_{z_i}^{z_f} d z \left(L_1 + \frac{12}{c} T(z) L_{-1} \right)} \right\} \< x | ,
\ee
where the $L_a$ are taken in the representation of equation (\ref{eq:RepresentationofL}).   We will be evaluating the matrix elements of these Wilson lines between primary states 
\be
\< h | W_h(z_f; z_i) | h \> = \left. P \left\{ e^{\int_{z_i}^{z_f} d z \left[ \partial_x + \frac{12}{c} T(z) \left( \frac{1}{2} x^2 \partial_{x} + h x \right) \right] } \right\} \frac{1}{x^{2h}} \right|_{x = 0} .
\ee
We emphasize that here $T(z)$ is the stress tensor \emph{operator}.  We can explicitly evaluate the Wilson line in $1/c$ perturbation, giving 
\be
\label{eq:AllOrdersWilsonLineFromStressTensor}
&& \< h| W_h(z; 0) | h \> \\ && \qquad = \left. \sum_{n=0}^{\infty} \left( \frac{6}{c} \right)^n \int_0^z dz_n \cdots \int_0^{z_2} dz_1 \left[ \prod_{i=1}^n T(z_i) \left( (x+z_{*i})^2 \partial_x + 2h(x+z_{*i}) \right) \right]  \frac{1}{(z+x)^{2 h}} \right|_{x = 0}, \nn
\ee
where $z_{*i} \equiv z - z_i$. This is a formula for the operator appearing in the OPE of $\CO_h(z) \CO_h(0)$.  We can evaluate it explicitly to write the OPE in terms of the stress tensor and its products.  
For example, to first non-trivial order 
\be
\< h | W_h(z_f;z_i)| h\> &=& \frac{1}{z_{fi}^{2h}} \left( 1 +\frac{1}{c}  \int_{z_i}^{z_f} dz f_1(z; z_i, z_f) T(z)  + \dots \right), \\
&& f_1(z; z_i, z_f) \equiv  \frac{12 h}{ z_{fi}}(z_f-z)(z-z_i) ,
\label{eq:vacblockf1kernel}
\ee
where $z_{fi} \equiv z_f - z_i$.
Instead of studying the OPE directly, we will evaluate the vacuum Virasoro block between two pairs of operators so that we can check our methods against known results \cite{Fitzpatrick:2015dlt, Chen:2016cms}. To avoid clutter, we will drop the bras and kets $\< h|, |h\>$ on $\< h | W_h | h\>$  in the following. The vacuum block can be written as a correlator of two Wilson lines 
\be
\label{eq:VacuumBlockasWilsonLineCorrelator}
\CV(z) = \left\< W_{h_1} (z;0)  W_{h_2} (\infty;1) \right\> \equiv \lim_{R\rightarrow \infty} \left\< W_{h_1}(z;0) R^{2h_2} W_{h_2}(R;1)\right\>,
\ee
as pictured in figure \ref{fig:PerturbativeVacuumBlock}.
Let us begin by evaluating the terms of order $\frac{1}{c}$, which correspond to stress tensor global conformal block, or `1-graviton exchange' in AdS$_3$: 
\be
\CV(z) &=& z^{-2h} \left[ 1 + \frac{144 h_1 h_2}{c^2 z} \int_0^z \int_1^\infty dz' dz'' \< T(z') T(z'')\> z'(z-z')(z''-1) + \dots \right] \nn\\
 &=& z^{-2h} \left( 1 + 2 \frac{h_1 h_2}{c} z^2 {}_2F_1(2,2,4,z) + \dots \right),
 \ee
 where we have used the two-point function $\< T(z') T(z'')\> = \frac{c}{2(z'-z'')^4}$ of the stress tensor. We recognize the second term $\frac{2 h_1 h_2}{c} z^2 {}_2 F_1(2,2,4,z)$ as 
the global conformal block associated with stress tensor exchange.

This first computation provides a nice check of the formalism.  But the effect that we have computed survives in the semiclassical limit,\footnote{Semiclassical effects  are terms in $\log \CV$  that are of order $c$ in the limit $c \to \infty$ with all $h_i / c$ fixed.} so it does not verify our methods at the quantum level.  However, we can use equation (\ref{eq:AllOrdersWilsonLineFromStressTensor}) to compute the vacuum Virasoro block to any order in $1/c$ perturbation theory.  The only additional complication arises from a need to regulate\footnote{Readers surprised by divergences in a Chern-Simons computation may  consult appendix \ref{app:ConnectionwithCovariantGauge}.
}   singular $T(z_i) T(z_j)$ OPEs when $z_i \to z_j$. 

Let us write the Wilson line at the operator level in a perturbative $1/c$ expansion as $W_h(z,0) = \sum_k W_h^{(k)}$ where the $k^{\rm th}$ term is proportional to $c^{-k}$.  At second order in $1/c$, the Wilson line $W_h(z;0)$ is the operator
\be
W_h^{(2)}(z_f; z_i) &=& \frac{1}{c^2z_{fi}^{2h}}\int_{z_i}^{z_f} dz_1 dz_2 T(z_1) T(z_2)  \left[ \frac{1}{2} f_1(z_1;z_f, z_i) f_1(z_2; z_f, z_i) + f_2(z_1, z_2; z_f, z_i) \right], \nn\\
f_2(z_1,z_2; z_f, z_i) &=& \frac{36 h}{z_{fi}^2} (z_f - {\rm max}(z_1, z_2) )^2({\rm min}(z_1, z_2) - z_i)^2 . 
\label{eq:vacblockf2kernel}
\ee
We have written the ``kernel'' above as a product of factors of the kernel $f_1$ at $\CO(1/c)$ plus a new term $f_2$. In section \ref{sec:subleadingfromPI}, we will demonstrate that this pattern continues to all orders, and the new term is always $\CO(h/c)$.   When we evaluate the vacuum block using equation (\ref{eq:VacuumBlockasWilsonLineCorrelator}), we will have terms of the form 
\be
\CV^{(2)} \supset \< W_{h_1}^{(2)} W_{h_2}^{(1)} \> + \< W_{h_1}^{(1)} W_{h_2}^{(2)} \> + \< W_{h_1}^{(2)} W_{h_2}^{(2)} \> 
\ee
which can contribute to the vacuum block at order $1/c^2$.  The first two terms involve 3-pt stress tensor corrleators $\<TTT\>$, while the last term involves the 4-pt correlator $\< TTTT \>$.  The integrals along the Wilson lines will diverge due to singular terms in the $T(z_2)T(z_1)$ OPEs.  As we discuss in detail in appendix \ref{app:Regulation}, one can choose a regulator that can be thought of at low orders as due to normal ordering, so that
\be
\< W_{h_1} W_{h_2} \>_{reg} = \< [W_{h_1}] \ [W_{h_2}] \>
\ee
that eliminates all OPE singularities within a given Wilson line.  
By definition, the vacuum expectation values of normal ordered products of $T$s vanish,
\be
\< [ T(z_1) \dots T(z_n) ] \> =0,
\ee
so the only terms that contribute to $\< [ W_{h_1} ] \ [ W_{h_2}] \>$ at leading order are terms where every $T$ in $W_{h_1}$ is contracted with a $T$ from $W_{h_2}$.  Manifestly, then, only those correlators with an equal number of $T(z_i)$ on each Wilson line survive and so the vacuum bock at order $1/c^2$ becomes
\be
\CV^{(2)} = \< [W_{h_1}^{(2)}] \ [W_{h_2}^{(2)}] \> .
\ee
According to the definition in appendix \ref{app:Regulation}, we have
\be
\< [T(z_1) T(z_1')] [ T(z_2) T(z_2')] \> &=&   \frac{c^2}{4z_{12}^4 z_{1'2'}^4} +  \frac{c^2}{4 z_{1 2'}^4 z_{1' 2}^4} + \CO(c) 
\ee
where we do not explicitly display terms at order $c$ and higher in the $1/c$ expansion.
To compute the conformal block at $\CO(1/c)$, we must keep the $\CO(c^2)$ ``disconnected'' part of the $\< [TT ] \ [TT ] \>$ correlator in the above line, drawn schematically in figure \ref{fig:PerturbativeVacuumBlock}, whereas we can discard the $\CO(c)$  ``connected'' part, which only contributes to the vacuum block at order $\frac{1}{c^3}$.
 
 Computing the two pairs of integrations for each Wilson line, we find
\be
\CV^{(2)} &=& 36 \frac{h_1^2 h_2 + h_1 h_2^2}{c^2} \left(   \frac{(z-2)z \log(1-z) + 2 (1-z) \log^2 (1-z)}{z^2}  - 4 \right) 
 \\
&& + \frac{12 h_1 h_2 }{c^2} \frac{\left(12 (z-2) z \text{Li}_2(z)+16 z^2+6 (z-1)^2 \log ^2(1-z)+(z-2) z \log (1-z)\right)}{z^2}  \nn 
\ee
which is in agreement with other calculations \cite{Fitzpatrick:2015dlt, Chen:2016cms} of the Virasoro vacuum block at order $1/c^2$.  This provides a quantum-level check of our formalism.

To compute at even higher orders in $1/c$ we simply apply equation (\ref{eq:AllOrdersWilsonLineFromStressTensor}) using appropriate multi-stress tensor correlators.  For example, at order $1/c^3$ we would need both the disconnected part of $\< [TTT] \ [TTT] \>$ and the next-to-leading-order connected part of $\< [TT] \ [TT] \>$.  It would be very interesting to check this $\frac{1}{c^3}$ computation using recent results derived by other methods \cite{Chen:2016cms}, especially since it is the first term involving a connected stress tensor correlator.  We have performed some partial checks, but only to low-orders in $z$. It may also be possible to use these results to provide a more natural derivation of the diagrammatic rules for the heavy-light vacuum block \cite{Fitzpatrick:2015foa}.

\subsection{General Virasoro Blocks at Order $\frac{1}{c}$}

In this section we demonstrate that Wilson line construction of Virasoro OPE blocks can be used to reproduce
the Virasoro blocks for 4-pt correlators with general intermediate states.  Specifically, we will compute the Wilson lines in $1/c$ perturbation theory and show that they match with known results \cite{Fitzpatrick:2015zha} 
and \cite{ZamolodchikovRecursion}. The starting point is to expand the Wilson lines in the Virasoro OPE block (\ref{eq:OPEfromWilsonLines}) order by order in $\frac{1}{c}$.  Writing $h$ for the external operator dimension and $h_p$ for the internal dimension, the first two orders are\footnote{When inserted in correlators, $T$ produces at most a factor of $c^{-1/2}$.  So this expansion is controlled at large $c$ even though it is an operator equation. }
\begin{equation}
\mathcal{V}_{p}=\frac{\Gamma(2h_{p})}{\Gamma(h_{p})^{2}}\frac{1}{z_{21}^{2h}}\int_{z_{1}}^{z_{2}}dz_{3}\left(\frac{z_{23}z_{31}}{z_{21}}\right)^{h_{p}-1}O_{p}(z_{3})\left[1+\frac{6}{c}\mathcal{T}(z_{2},z_{1},z_{3})+\mathcal{O}\left(\frac{T^{2}}{c^{2}}\right)\right]\label{eq:VOPEto1/c},
\end{equation}
where 
\begin{equation}
\mathcal{T}=2h\mathcal{T}_{1}(z_{2},z_{1})+(h_{3}-1)\mathcal{T}_{2}(z_{2},z_{1},z_{3}),
\label{eq:1/cOPEBlockT}
\end{equation}
and
\begin{equation}
\mathcal{T}_{1}(z_{2},z_{1})=\frac{1}{z_{21}}\int_{z_{1}}^{z_{2}}dw(z_{2}-w)(w-z_{1})T(w), \label{eq:T1}
\end{equation}
\begin{equation}
\mathcal{T}_{2}(z_{2},z_{1},z_{3})=\frac{z_{23}}{z_{21}z_{31}}\int_{z_{1}}^{z_{3}}dw(w-z_{1})^{2}T(w)+\frac{z_{31}}{z_{21}z_{23}}\int_{z_{3}}^{z_{2}}dw(z_{2}-w)^{2}T(w) .
\end{equation}
At leading order equation (\ref{eq:VOPEto1/c}) is just the global OPE
block for $O_{p}$. At each order, an infinite number of global primaries
built from $T^{n}O_{p}$ type-operators will generically be included, with their coefficient
determined by the Virasoro symmetry. For example, the $n=1$ order
we explicitly displayed already resums an infinite tower of global
OPE blocks of $L_{-n}O_{p}$ with $n\ge2.$\footnote{For each $L_{-n}O_{p}$, one needs to substract the descendant pieces
to construct a global conformal primary. For example, $(L_{-2}-\frac{3}{2(2h_{p}+1)}L_{-1}^{2})O_{p}$
is the primary component within $L_{-2}O_{p}$. }

Compared to the large amount of data involved in the organization of these Virasoro descendants, the Wilson line construction points to a remarkably succinct representation of the Virasoro OPE block (\ref{eq:OPEfromWilsonLines}) (\ref{eq:VOPEto1/c}). However, this representation contains UV divergences as the operators $T$ and $O_{p}$ approaches each other on the Wilson line, which need to be consistently regularized to be useful in computing correlation functions. Luckily, the representation of the Virasoro OPE block as a sum over global OPE blocks implies the existence of an unique and well defined regularization scheme. At low orders, this scheme coincides with that  detailed in Appendix \ref{app:Regulation}.

We will put (\ref{eq:VOPEto1/c}) to use and explicitly demonstrate
the power of the Virasoro OPE block. In particular, we use a pair
of these Virasoro OPE blocks to compute the $\frac{1}{c}$ expansion
of the Virasoro block of $O_{p}$ in the 4-point function $\langle O_{1}(0)O_{1}(z)O_{2}(1)O_{2}(\infty)\rangle$.
At leading and next to leading order, this Virasoro block takes the form
\begin{equation}
\mathcal{V}_{p}=g(h_{p},z)+\frac{h_{1}h_{2}}{c}f_{a}(h_{p},z)+\frac{h_{1}}{c}f_{b}(h_{p},z)+\frac{h_{2}}{c}f_{b}(h_{p},z)+\frac{1}{c}f_{c}(h_{p},z)+\mathcal{O}\left(\frac{1}{c^{2}}\right) .
\end{equation}
The terms $f_{a}$ and $f_{b}$ are determined by the semi-classical
Virasoro block\footnote{The $\frac{h_{2}}{c}$ piece directly appears in the semi-classical
result. The $h_{1}/c$ term is related to the $h_{2}/c$ term
by permutation symmetry, which is $z_{1},z_{2}\leftrightarrow z_{3},z_{4}$
and leaves $z$ unchanged. } computed in \cite{Fitzpatrick:2015zha}, while the $f_{c}$ piece is a quantum correction to the semiclassical
result that has never been computed in closed form. 

Using our Wilson line formalism, at $\frac{1}{c}$ order of the Virasoro
block is: 
\begin{equation}
V_{p}|_{\frac{1}{c}}=\frac{36}{c^{2}}\left(\frac{\Gamma(2h_{p})}{\Gamma(h_{p})^{2}}\right)^{2}\int_{z_{1}}^{z_{2}}dz_{5}\int_{z_{3}}^{z_{4}}dz_{6}\frac{\langle O_{p}(z_{5})\mathcal{T}(z_{2},z_{1},z_{5})O_{p}(z_{6})\mathcal{T}(z_{4},z_{3},z_{6})\rangle}{z_{21}^{2h_{1}+h_{p}}z_{25}^{1-h_{p}}z_{51}^{1-h_{p}}z_{43}^{2h_{2}+h_{p}}z_{46}^{1-h_{p}}z_{63}^{1-h_{p}}},\label{eq:1/cBlockForm}
\end{equation}
where the regulator requires that we do not include self-contractions, meaning that
$\langle O_{p}\mathcal{T}O_{p}\mathcal{T}\rangle\rightarrow\langle O_{p}O_{p}\rangle\langle\mathcal{T}\mathcal{T}\rangle$ at this order. 

We first compute $f_{a}$ using the $\mathcal{T}_{1}$ term in (\ref{eq:1/cOPEBlockT}).
The OPE block is then remarkably simple. We find
\begin{equation}
\mathcal{V}_{p}|_{\frac{h_{1}h_{2}}{c}}=\frac{1}{z_{21}^{2h}}\int_{z_{1}}^{z_{2}}dz_{3}\left(\frac{z_{23}z_{31}}{z_{21}}\right)^{h_{p}-1}O_{p}(z_{3})\int_{z_{1}}^{z_{2}}dw\frac{(z_{2}-w)(w-z_{1})}{z_{21}}T(w).
\end{equation}
This is simply the product of the global OPE block of $O_{p}$ and
$T$. With the normal ordering of operators, this implies
\begin{equation}
f_{a}=g(h_{p},z)\frac{\Gamma(2)^{2}}{\Gamma(4)}g(2,z)=-12z^{h_{p}-1-2h_{1}}F_{21}(h_{p},h_{p},2h_{p},z)(2z+(2-z)\log(1-z)).
\end{equation}
This precisely agrees with the known result \cite{Fitzpatrick:2015zha}. 

The $f_{b}$ piece comes from the mixed term $\langle\mathcal{T}_{2}(z_{2},z_{1},z_{5})\mathcal{T}_{1}(z_{4},z_{3})\rangle$
in (\ref{eq:1/cBlockForm}). The calculation is equally straightforward.
The result is
\begin{equation}
f_{b}=12h_{p}z^{h_{p}-2h_{1}}\left[\left(\frac{(1-z)\log(1-z)}{z}+1\right)F\left(h_{p},h_{p};2h_{p};z\right)+\frac{\log(1-z)}{2}F\left(h_{p},h_{p};2h_{p}+1;z\right)\right].
\end{equation}
This also agrees with \cite{Fitzpatrick:2015zha}. 

The $f_{c}$ function comes from the mixed term $\langle\mathcal{T}_{2}(z_{2},z_{1},z_{5})\mathcal{T}_{2}(z_{4},z_{3},z_{6})\rangle$ in equation (\ref{eq:1/cBlockForm}). The calculation is more complicated due
to the explicit dependence of this correlator on $z_{5,6}$. We have
not computed it in closed form. Instead, we obtain the first few orders
in the small $z$ expansion: 
\begin{eqnarray}
f_{c} & = & \frac{h_{p}^{2}(h_{p}-1)^{2}}{2\left(2h_{p}+1\right)^{2}}\left[z^{2}+\frac{h_{p}+2}{2}z^{3}+\frac{\left(h_{p}+3\right)\left(h_{p}\left(10h_{p}\left(2h_{p}+11\right)+191\right)+108\right)}{40\left(2h_{p}+3\right){}^{2}}z^{4}\right.\nonumber \\
 &  & +\left.\frac{\left(h_{p}+3\right)\left(h_{p}+4\right)\left(h_{p}\left(10h_{p}\left(2h_{p}+13\right)+243\right)+144\right)}{240\left(2h_{p}+3\right){}^{2}}z^{5}+\mathcal{O}(z^{6})\right].
 \label{eq:1/cBlockExp}
\end{eqnarray}
This agrees with the small $z$ expansion of $\mathcal{V}_{p}$ obtained from  the Zamolodchikov recursion relations \cite{ZamolodchikovRecursion}. This provides a highly non-trivial check that the Wilson line formalism works at the quantum level.

\section{Path Integral Formalism and  Bulk Reconstruction}
\label{sec:PathIntegralFormalismandReconstruction}

The sl$(2)$ Wilson lines can be interpreted as operators that propagate $x$-space wavefunctions along a path in physical spacetime.    Taking this idea seriously leads to a new presentation of Wilson lines in terms of a path-integral, which we derive in section \ref{sec:DerivationPathIntegral}.   The path integral formalism makes the semiclassical limit of the Wilson lines manifest.  We develop this point in section \ref{sec:HeavyLightandUniformizing}, showing how one can immediately obtain heavy-light Virasoro blocks and a variety of other correlators in a heavy background.  Thus  the Wilson lines automatically reconstruct geometry in a background-independent way.

\subsection{Derivation of a Path Integral Formula}
\label{sec:DerivationPathIntegral}

In this section, we will derive a simpler expression for the Wilson line by writing it as a path integral. As we saw in equation (\ref{eq:xDefinitionWilsonLine}), the Wilson lines act as evolution operators that describe how states evolve as a function of $z$, $W_h(z_f, x_f; z_i, x_i) = \< x_f; z_f | x_i; z_i\>$.  The formula (\ref{eq:WilsonLineHamiltonian}) for the ``Hamiltonian'' $H$ simplifies near the boundary, per equation (\ref{eq:WilsonLineNearBoundary}), to become
\be
-i H_W(z) &\equiv& i P + \frac{6 T(z)}{c} \left( i X^2 P + 2 h X\right),
\ee
The form of $H_W$ makes it straightforward to write $W$ as a path integral\footnote{Note that this is distinct from a path integral description used in previous work \cite{Ammon:2013hba}, as we are not including a dynamical particle moving along the Wilson line.}
\begin{equation}
W(z_f, x_f; z_i; x_i) = \int {\cal D}p(z) \int_{x(z_i)=x_i \atop x(z_f) = x_f } {\cal D}x(z) e^{ \int_{z_i}^{z_f} dz \left( i p \left(  \frac{dx}{dz} + 1 + \frac{6 T(z)}{c} x^2 \right) + \frac{6 T(z)}{c} 2h x \right)  }.
\end{equation}
The fact that $p(z)$ appears linearly means that it simply acts as a Lagrange multiplier imposing the constraint
\be
-x'(z) &=& 1 + \frac{6 T(z)}{c} x^2(z).
\label{eq:xTconstr}
\ee
The integral ${\cal D}x(z)$ then becomes trivial since it is localized by the $\delta$ function for this condition.\footnote{The Jacobian factor for this $\delta$ function is trivial; one way to see this is to discretize $x(z) \rightarrow x_i$ and recursively evaluate the integrals ${\cal D}x(z) \rightarrow \prod_j \int d x_j$ starting with $x_f=x(z_f)$ first, so at each step the $\delta$ function appears in the integral as $\int dx_{j+1} \delta(x_j - x_{j+1} + \epsilon (1+ \frac{6 T(z_j)}{c} x_j^2)) = 1$, with $\epsilon$ being the discretization length.}  Because $T(z)$ is an operator, this constraint effectively promotes $x(z)$ to an operator as well.  To be explicit, one can solve this equation for $x(z)$ order by order in $1/c$, in which case one obtains a representation for $x(z)$ as sums over integrals of products of $T(z)$.  Labeling this solution, subject to the boundary condition $x_T(z_f) = x_f$, as ``$x_T(z)$'', we can write the Wilson line as
\be
\boxed{W(z_f, x_f; z_i, x_i) = \left( e^{\int_{z_i}^{z_f} dz \frac{12 h T(z)}{c} x_T(z)}\right) \delta(x_i - x_T(z_i)) .}
\label{eq:unitaryevolutionop}
\ee
In the limit $c\rightarrow \infty$ with other parameters fixed, the evolution operator becomes trivial.  We can easily solve the constraint equation:
\be
x_T(z) = x_f - (z-z_f), 
\ee
and therefore the Wilson line reduces to
\be
\lim_{c \rightarrow \infty} W(z_f, x_f; z_i, x_i) &=& \delta(x_{fi} + z_{fi}) ,
\ee
which is just (\ref{eq:InfCEvol}) in the $x$ basis. 
At general $c$, a Wilson line with primary endpoints can be written in the compact form 
\begin{equation}
\< h| W( z_f, z_i)| h\> =
\left( e^{\int_{z_i}^{z_f} dz \frac{12 T(z)}{c} x_T(z)}\frac{1}{x_T(z_i)^2} \right)^h , 
 \quad -x'_T(z) = 1 + \frac{6 T(z)}{c} x_T^2(z), \quad x_T(z_f) =0.
 \label{eq:PathIntegralWilsonLine}
\end{equation}
where the function $x_T(z)$ is defined by this differential equation.

\subsection{Heavy-Light Limit and Uniformizing $w$-Coordinates}
\label{sec:HeavyLightandUniformizing}

As our first application of equation (\ref{eq:PathIntegralWilsonLine}), we will consider how the Wilson line behaves in the background created by a single heavy state with dimension of $\CO(c)$.  In \cite{Fitzpatrick:2015zha}, it was found that in the semi-classical limit $c\rightarrow \infty$, all insertions of the semiclassical stress tensor could be absorbed into a change of coordinates $z \rightarrow w(z)$, allowing a simple computation of Virasoro blocks.  We will show how this follows automatically from the Wilson line prescription formulated as a path integral.  Then we will use it to derive heavy-light vacuum blocks, the correlator of three light operators in a heavy operator background, and the general heavy-light Virasoro blocks \cite{Fitzpatrick:2015zha}, all in an arbitrary background for the stress tensor $T(z)$.  

First, we review some previous results.  If we look at the Wilson line $W_L$ in the heavy state $|\Psi\>$, in the semiclassical limit we can treat the stress tensor as a $c$-number function given by its expectation value,
\be
T_\Psi(z) &\equiv& \frac{\< \Psi | T(z) | \Psi\>}{\< \Psi | \Psi\>} .
\ee
If $|\Psi\>$ is a primary state of weight $h_H$ inserted at the origin, then $T_\Psi(z) = \frac{h_H}{z^2}$, though we will not need to restrict to this case.  For any $T_\Psi(z)$, the uniformizing ``$w(z)$'' coordinates are just those coordinates in which the expectation value of the stress tensor vanishes (or at least is $\CO(1)$ rather than $\CO(c)$) due to a cancellation with the Weyl anomaly:
\be
T_\Psi(w) &=& (w'(z))^{-2} \left( T_\Psi(z) - \frac{c}{12} S(w,z) \right) , 
\ee
where $S(w,z)$ is the Schwarzian derivative.  Demanding $T_\Psi(w)=0$ implies a third-order differential equation for $w(z)$:
\be
 S(w,z) = \frac{w'''(z)}{w'(z)} - \frac{3}{2} \left( \frac{w''(z)}{w'(z)} \right)^2  = \frac{12}{c}T_\Psi(z).  
\label{eq:UniformizingCoords}
 \ee
 The resulting expression for the vacuum Virasoro block is particularly simple 
 \be
 \CV(z_f, z_i) &=& \left( \frac{w'(z_f) w'(z_i) }{(w(z_f) - w(z_i))^2} \right)^h.
 \label{eq:SCVacBlock}
 \ee
In the specific case where the background value $T_{\psi}(z) = \frac{h_H}{(1-z)^2}$ comes from a pair of heavy primary operators, we obtain the semiclassical heavy-light Virasoro vacuum block with $w(z) = 1 - (1-z)^\alpha$ and $\alpha = \sqrt{1 - \frac{24 h_H}{c}}$ as discussed in \cite{Fitzpatrick:2015zha}.  

However the fundamental idea is much more general -- any background $T_\Psi$ from heavy operator sources can be absorbed into the background metric by transforming to the uniformizing coordinate $w(z)$.  Then the contribution to the two point function of light operators with $h_L \ll c$ from  the exchange of Virasoro vacuum descendants with the background will take the form of equation (\ref{eq:SCVacBlock}).  We can obtain very general results by evaluating the Virasoro OPE block in such a background.
 
 \begin{figure}[t!]
\begin{center}
\includegraphics[width=0.95\textwidth]{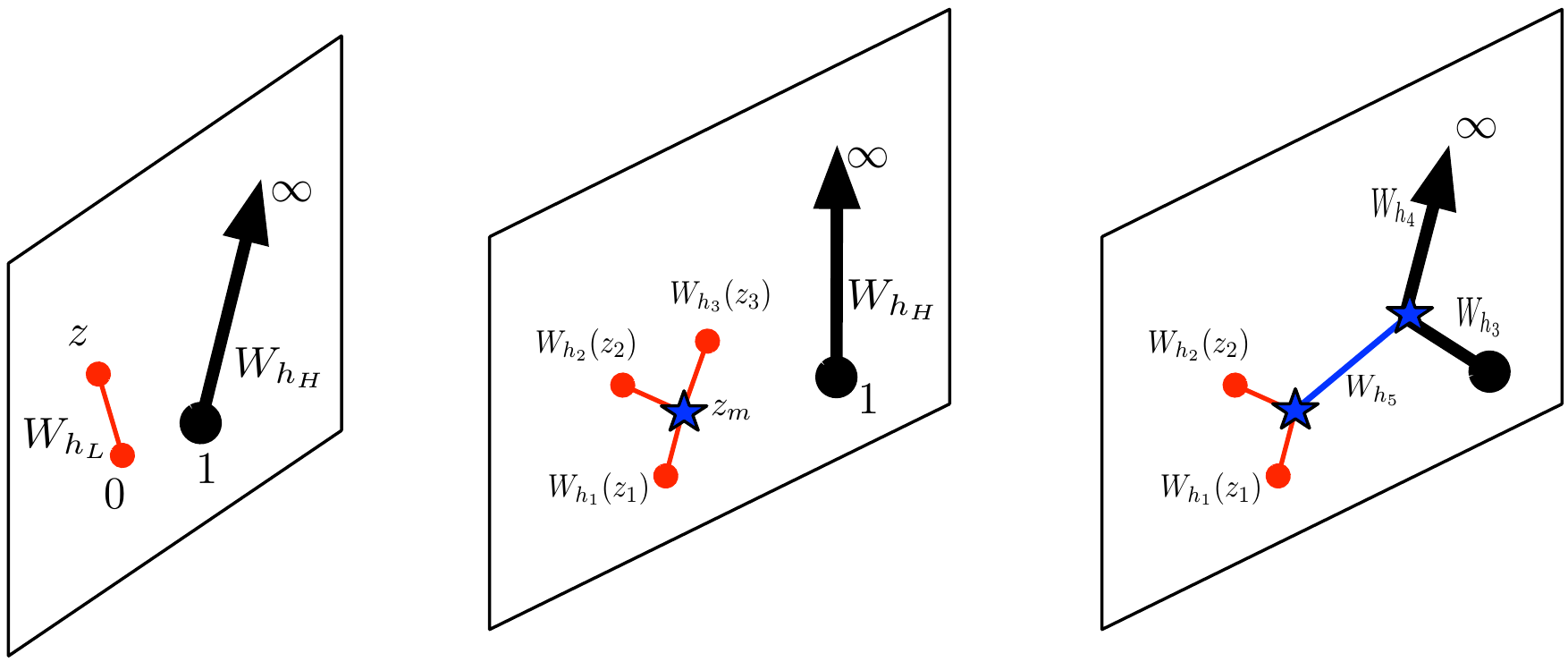}
\caption{This figure indicates configurations of Wilson lines for computing the heavy-light Virasoro vacuum block, a light operator 3-pt correlator in a heavy operator background, and a general non-vacuum Virasoro block.  Dots indicate boundary points, whereas stars and the Wilson line trajectories themselves are free to float off into the bulk.  The thick black lines suggest `heavy' Wilson lines with $h_H \propto c$ in the large $c$ limit. }
\label{fig:HeavyLightVacuum3ptandGeneral}
\end{center}
\end{figure} 

\subsubsection{Two-point Function Vacuum Block}
 
To see how these uniformizing $w$-coordinates are connected to the Wilson line, note that if we identify $x_T(z)$ in terms of $w(z)$ through
\be
\frac{1}{x_T(z)} &=& \frac{w''(z)}{2w'(z)}- \frac{w'(z)}{w(z)+C},
\ee
then $x_T(z)$ automatically satisfies the differential constraint equation (\ref{eq:xTconstr}) for any value of $C$; the boundary condition $x(z_f)=0$ corresponds to the choice 
\be
C= - w(z_f).
\ee
   The fact that $x_T$ satisfies a first order differential equation whereas $w$ satisfies a third order equation reflects the fact that the $w$ equation of motion is invariant under both a scaling $w\rightarrow \lambda w$ and a shift $w \rightarrow w+c$.\footnote{In the language of   
   \cite{Turiaci:2016cvo,Dorn:2006ys,Teschner:1995dt}, $\frac{1}{x_T} \sim \varphi'$, where $\varphi$ is treated as a periodic free field.  Written in terms of $\varphi$, our results are very similar to, but not manifestly the same as those of Guica \cite{Guica:2016pid}.}
   
Next, substitute this solution into our formula (\ref{eq:PathIntegralWilsonLine}) for a light operator (with $h \ll c$) Wilson line.  As a consequence of equation (\ref{eq:xTconstr}) we have
   \begin{equation}
   \int_{z_i}^{z_f}  dz \frac{6T(z)}{c} x_T(z) = - \int_{z_i}^{z_f} dz \frac{x'_T(z)+1}{x_T(z)}  = - \left[ \log(x_T(z)) + \log\left(\frac{(w'(z))^{1/2}}{w(z) +C}\right)\right]_{z_i}^{z_f} .
   \end{equation}
So we see that equation (\ref{eq:PathIntegralWilsonLine}) can be written entirely in terms of the uniformizing coordinate $w(z)$ and its derivatives. We therefore find
\begin{equation}
\< h| W( z_f, z_i)| h\> = \lim_{C \rightarrow -w(z_f)} \left( e^{-2\int_{z_i}^{z_f} dz \frac{x_T'(z) +1}{ x_T(z)} }\frac{1}{x_T(z_i)^2} \right)^h  
= \left(\frac{w'(z_f) w'(z_i)}{(w(z_f) - w(z_i))^2}\right)^h,
\label{eq:vacuumblockfromWL}
\end{equation}
exactly reproducing (\ref{eq:SCVacBlock}) for an arbitrary heavy background.   The case where this computes the heavy-light vacuum block is pictured on the left in figure  \ref{fig:HeavyLightVacuum3ptandGeneral}.

\subsubsection{Three-point Function Vacuum Block}

The  simple result (\ref{eq:vacuumblockfromWL}) for the two-point function generalizes to describe the case of a network of Wilson lines with $h_i \ll c$.  Let us consider a network of three Wilson lines beginning at $z_1, z_2, z_3$ and meeting at a gauge-invariant vertex at $z_m$, as pictured in the center of figure \ref{fig:HeavyLightVacuum3ptandGeneral}.  This is
\be
W_{123} = \int dx_1 dx_2 dx_3  W_{h_1}(z_1, z_m) W_{h_2}(z_2, z_m) W_{h_3}(z_3, z_m) f_{12 3}(x_1, x_2, x_3) 
\ee
where $f_{123}$ is a sl$(2)$ invariant vertex.  For convenience, we will define a function
\be
X(z, a) \equiv  \left(  \frac{w'(z)}{w(z) - w(a)} - \frac{w''(z)}{2w'(z)}  \right)^{-1}
\ee
that automatically solves the equation of motion for $x_T(z)$ with a specific boundary condition, i.e.
\be
\Big[ x_T(z) \Big]_{x_T(a)=0} = X(z,a).
\ee
Each of the three Wilson lines will supply  a factor of
\be
e^{h_i \int_{z_m}^{z_i}  dz \frac{12T(z)}{c} x_T(z)}  = 
\left( \frac{X(z_m, z_i) \sqrt{w'(z_i) w'(z_m)} }{w(z_m)- w(z_i)} \right)^{2h_i}
\label{eq:WilsonLineExpSimp}
\ee
Including delta functions that choose primary states at the $z_i$, the bulk vertex is
\be
f_{12 3}(x_1, x_2, x_3) = \frac{\delta(x_1 - X(z_m,z_1)) \delta(x_2 - X(z_m, z_2)) \delta(x_3 - X(z_m,z_3))}{x_{12}^{h_1 + h_2 - h_3 } x_{23}^{h_2 + h_3 - h_1} x_{31}^{h_3 + h_1 - h_2} }
\ee
The powers of $X$ from the Wilson lines combine with the powers of $x_{ij}$ from the bulk vertex via
\be
\frac{X(z_m, z_i) X(z_m, z_j)}{X(z_m, z_i) - X(z_m, z_j)} 
=
\frac{(w(z_m) - w(z_i)) (w(z_m) - w(z_j))}{(w(z_j) - w(z_i)) w'(z_m)}
\label{eq:XfromWilsonLinesIdentity}
\ee
which can be seen easily by noting that the variable $\frac{1}{X}$ is simpler than $X$ itself.  It is particularly important that the dependence on the intermediate point $z_m$ has simplified.  Combining all three Wilson lines, we find the simple final result
\be
\< \Psi | W_{123} | \Psi \> = \frac{(w'_1)^{h_1} (w'_2)^{h_2} (w'_3)^{h_3}}{w_{12}^{h_1 + h_2 - h_3 } w_{23}^{h_2 + h_3 - h_1} w_{31}^{h_3 + h_1 - h_2} }
\ee
where $w_{ij} = w(z_i) - w(z_j)$ and $w'_i = w'(z_i)$.
This result precisely agrees with what we would expect for a 3-pt CFT$_2$ correlator transformed to the uniformizing $w$-coordinate background.  Notice that all dependence on the intermediate point $z_m$ has dropped out of this final expression, which depends only on $w(z)$, and the locations of the points $z_i$ and their corresponding holomorphic dimensions $h_i$.

\subsubsection{Heavy-Light Non-Vacuum Block}

The previous two examples -- the two-point function vacuum block and the three-point function vacuum block -- exchanged only the vacuum Virasoro representation between the Wilson line and the other states in the correlation function. As a result, the only aspect of the background state that mattered was the expectation value it gave to the stress tensor $\< T(z)\>$.  To consider non-vacuum blocks, we also have to include information about how the primary operator in the exchanged representation responds to the background state.  That is, the non-vacuum block can be thought of as simply the expectation value of the Virasoro OPE block in the presence of two primary operators:\footnote{The background does not necessarily have to be created by two primary operators; all that is required is that there exists a coordinate system where $\frac{\<T\>}{c}$ and $\< \CO_p\>$ have finite $c\rightarrow \infty$ limits.} 
\be
\< \CO_{H_1}(\infty) \CO_{H_2}(1) \CO_{L}(z_f) \CO_{L}(z_i)\>  = \< \CO_{H_1}(\infty) \CO_{H_2}(1) : W_L(z_f; z_i) :\> .
\label{eq:BlockFromWVev}
\ee

First, we need to understand how to evaluate the expectation value of the OPE block using purely CFT arguments, so that we know what to compare to when we evaluate our Wilson lines.    In the semi-classical limit, \cite{Fitzpatrick:2015zha} showed that the block reduces to the insertion of a projection operator involving only the generators $\CL_{-1}$ of translations in $w$ coordinates:
\be
\CV_{h_p} = \< \CO_{H_1}(\infty) \CO_{H_2}(1)   \left( \sum_{k=0}^\infty \frac{\CL_{-1}^k | h_p \> \< h_p | \CL_1^k}{\< h_p | \CL_1^k \CL_{-1}^k | h_p \> }  \right) \CO_L(w_f) \CO_L(w_i) \> . 
\ee
The matrix elements $\< \CO_{H_1} \CO_{H_2} \CL_{-1}^k | h_p\>$ can all be read off from the series expansion of the correlator $\< \CO_{H_1} \CO_{H_2} \CO_{h_p}(w)\>$, which one can express as the expectation value of $\CO_{h_p}$ in the background of the heavy states.  Let $f(w)$ be this correlator:
\be
f(w) \equiv \< \CO_{H_1}(\infty) \CO_{H_2}(1) \CO_{h_p}(w) \> = \< H_1 | \CO_{h_p}(1-w) | H_2\>.
\ee
The exact relation between $f(w)$ and the matrix elements we need is simply
\be
\< \CO_{H_1} \CO_{H_2} \CL_{-1}^k | h_p\> = f^{(k)}(0).
\ee
Next, we can trade the sum over $k$ for an integral, using the following identity:
\be
\frac{\< h_p | \CL_1^k \CO_{L}(w_f) \CO_{L}(w_i) \>}{\< h|  \CL_{+1}^k \CL_{-1}^k |h \>} =\frac{(w'(z_f)w'(z_i))^{h_L}}{w_{fi}^{2h_L+h_p-1}} \frac{\Gamma(2h_p)}{\Gamma^2(h_p)} \int_{w_i}^{w_f} dw (w-w_i)^{h_p-1} (w_f-w)^{h_p-1} \frac{w^k}{k!}. \nn\\
 \ee
Combining this identity with the projector, we see that the block can be written as
\be
\CV_{h_p} = \frac{(w'(z_f)w'(z_i))^{h_L}}{w_{fi}^{2h_L+h_p-1}} \frac{\Gamma(2h_p)}{\Gamma^2(h_p)} \int_{w_i}^{w_f} dw (w-w_i)^{h_p-1} (w_f-w)^{h_p-1} f(w).
\label{eq:SCNonvacBlock}
\ee

This expression is the semi-classical limit of the conformal block as derived from the CFT.  We will now see how it is reproduced from the Wilson line prescription for the OPE block.  Putting together (\ref{eq:OPEfromWilsonLines}),  (\ref{eq:GenOPEBlock1}), and (\ref{eq:BlockFromWVev}), the Wilson line prescription for a general Virasoro block takes the form
\be
\CV^{(\rm WL)}_{h_p} &=& \int_{z_i}^{z_f}  dz_3 \,  \< \CO_{H_1}(\infty) \CO_{H_2}(1) \CO_{h_p}(z_3)\> \, F_{12,3} (z_f, z_i; z_3) \nn \\
 &=& \left. \int_{z_i}^{z_f} dz_3 \right[ \< \CO_{H_1}(\infty) \CO_{H_2}(1) \CO_{h_p}(z_3)\>  
 \nn \\
&& \left. \times \int dx_1 dx_2 W_{h_L} (z_i; 0; z_3, z_1) W_{h_L}(z_f; 0 ; z_3, x_2) f_{12 \tilde p} (x_1, x_2, 0) \right]. 
\ee
The three-point function $\< \CO_{H_1} (\infty) \CO_{H_2}(1) \CO_{h_p}(z_3)\>$ is related to $f(w)$ according to the usual transformation rule of primary operators, $\CO_{h_p}(z) = \CO_{h_p}(w(z)) (w'(z))^{h_p}$.  After inserting the expression (\ref{eq:unitaryevolutionop}) for the Wilson lines, the integrals over $x_1, x_2$ are performed trivially thanks to the $\delta$ functions.   We also use the simple expression (\ref{eq:WilsonLineExpSimp}) for the exponentials to obtain
\be
\CV^{(\rm WL)}_{h_p} &=& \int_{z_i}^{z_f} dz_3 
\left( \frac{X(z_3, z_i) \sqrt{w'(z_i) w'(z_3)} }{w(z_3)- w(z_i)} \right)^{2h_L}
\left( \frac{X(z_3, z_f) \sqrt{w'(z_f) w'(z_3)} }{w(z_3)- w(z_f)} \right)^{2h_L}\nn\\
&&  f_{12\tilde{p}}(X(z_3, z_i), X(z_3, z_f), 0)  f(w(z_3))   (w'(z_3))^{h_p}. 
\ee
Finally, substituting the form (\ref{eq:shadow3point}) for the shadow three-point function and using the identity  (\ref{eq:XfromWilsonLinesIdentity}), many pleasing cancellations occur and leave behind the following expression
\be
\CV_{h_p}^{(\rm WL)} &=& \frac{(w'(z_f)w'(z_i))^{h_L}}{w_{fi}^{2h_L+h_p-1}} \int_{w_i}^{w_f} dw (w-w_i)^{h_p-1} (w_f-w)^{h_p-1} f(w),
\ee
in exact agreement\footnote{A similar analysis with different weights $h_{L_1} \ne h_{L_2}$ for the two light operators $\CO_{L_1}$ and $\CO_{L_2}$ is straightforward, and verifies the non-vacuum block from the Wilson line in this more general case.} with the semi-classical formula (\ref{eq:SCNonvacBlock}), up to the constant normalization factor $ \frac{\Gamma(2h_p)}{\Gamma^2(h_p)}$.

\subsection{Subleading $1/c$ Expansion}
\label{sec:subleadingfromPI}

The path integral representation of the Wilson line can also be used to  streamline the computation of the $1/c$ expansion of the blocks.  In this subsection, we will work this out for the vacuum Virasoro OPE block. To begin,  define the variable $y=1/x$, so that
\be
y'(z) &=& - y^2(z) - \frac{6 T(z)}{c} .
\ee
The Wilson line in terms of $y$ also simplifies:
\be
\< h| W(z_f; z_i)| h\> &=& \left( z_{fi}^{-2h} e^{-2 h \int_{z_i}^{z_f} dz ( y(z)  - \frac{1}{z-z_i})} \right).
\label{eq:WilsonLineReg}
\ee
Our strategy will be to solve the equation for $y$ order by order in $1/c$:
\be
y(z) &=& \sum_{n=0}^\infty c^{-n} y_n(z). 
\ee
At leading order, the solution is just  $y_0(z) = (z-z_i)^{-1}$, and the differential equation for the first order perturbation is simply
\be
y_1'(z) &=& - 2 y_0(z) y_1(z) - \frac{ 6 T(z)}{c}.
\ee
This has solution
\be
y_1(z) &=& - \frac{1}{(z-z_i)^2} \int_{z_i}^z dz' (z'-z_i)^2 6 T(z').
\ee
At higher orders, the source term $6 T(z)$ doesn't contribute.  Expanding out $y^2$ into its series expansion and matching terms of the same order, we find the recursion relation
\be
y_s'(z) &=& -2 y_0(z) y_s(z) - \sum_{n=1}^{s-1} y_n(z) y_{s-n}(z), \qquad (s>1) .
\ee
We can write the solution to this as a formal integral:
\be
y_s(z) &=& -\frac{1}{(z-z_i)^2} \int_{z_i}^z dz' (z'-z_i)^2 \sum_{n=1}^{s-1} y_n(z') y_{s-n}(z').
\ee
The above recursion formula is an algorithm for $y_s(z)$ at any order, and at each order generates an additional $T(z_i)$ and an integral $dz_i$ over its position.  Substituting them back into the exponent in (\ref{eq:WilsonLineReg}), we obtain an expansion of the form
\be
\log \Big( \< h| W(z_f; z_i)| h\>  z_{fi}^{2h} \Big) &=&
 \int_{z_i}^{z_f} dz_1 f_1(z_1) \frac{T(z_1)}{c} + \int_{z_i}^{z_f} dz_1 dz_2 f_2(z_1, z_2) \frac{ T(z_1) T(z_2)}{c^2} + \dots . \nn\\
\ee
Carrying out this procedure up to $\CO(1/c^2)$ reproduces the kernels $f_1(z)$ and $f_2(z_1, z_2)$ previously obtained in equations (\ref{eq:vacblockf1kernel}) and (\ref{eq:vacblockf2kernel}), respectively:
\be
f_1(z) &=& \frac{12 h}{ z_{fi}}(z_f-z)(z-z_i), \nn\\
f_2(z_1, z_2) &=&  \frac{36 h}{z_{fi}^2} (z_f - {\rm max}(z_1, z_2) )^2({\rm min}(z_1, z_2) - z_i)^2 . 
\ee
The form of (\ref{eq:WilsonLineReg}) makes it manifest that every kernel $f_n(z_1, \dots, z_n)$ in the exponent is proportional to exactly one power of $h$.  The multiple powers of $h$ that appear in the block arise from expanding the exponential $e^{h f_i} \sim 1+ h f_i + \dots$, and consequently at each order in $1/c$ the only really new term to compute is the linear in $h$ term, as the others can be read off from lower orders.  This explains the structure of the second-order part of the Wilson line we saw in equation (\ref{eq:vacblockf2kernel}).

\section{Discussion}
\label{sec:Discussion}

The main result of this paper is an exact expression for the Virasoro OPE blocks.  These objects resum all Virasoro descendants of a single primary operator $\CO_3$ in the OPE $\CO_1(z) \CO_2(0)$, and are written in terms of Wilson lines in the Chern-Simons formulation of AdS$_3$ gravity.   They can be used to efficiently compute the Virasoro blocks for $n$-point correlators.  In general dimensions, expressions that encapsulate the (global) conformal descendants have proven useful for organizing and studying the contributions of conformal irreps \cite{SimmonsDuffin:2012uy, Czech:2016xec, Hartman:2016lgu}, and we expect that our results may be of similar use.  

However, our primary motivation was to construct background-independent operators for use in exploring bulk physics in future work.  One remarkable property of the Virasoro OPE blocks is that they encode all non-linear effects from the dressing of local primary operators by products of stress tensors, in an arbitrary background.   Translated into AdS$_3$,  the Virasoro OPE blocks fully incorporate effects from the quantum gravitational field.

Thus the Virasoro OPE blocks are general, manifestly state-independent operators.  When they are inserted in a specific background state, they automatically piece together the appropriate coordinate system that uniformizes the boundary metric.  
This is made possible by the fact that  AdS$_3$ gravity is in some sense `kinematic', i.e. it is controlled by the Virasoro symmetry of the theory.  While we have focused here on OPE blocks, which can be interpreted as integrals of bulk operators along geodesics \cite{Hijano:2015qja,Hijano:2015zsa}, there is a close connection between these objects and local operators in the bulk \cite{Czech:2016xec, Guica:2016pid}.  Moreover, as exact operators, the Virasoro OPE blocks will incorporate quantum corrections to the semi-classical geometry, and once lifted into the bulk, they should be sensitive to its non-perturbative demise.  In other words, we expect that bulk geometry is only an approximate, emergent feature of CFT, and ideally a formalism for its description will predict its own range of validity.  The Virasoro OPE blocks and Chern-Simons Wilson lines appear to be the kinematic ingredients we need to construct  this formalism.

The Chern-Simons description we have used may also shed light on the subleading ``saddle'' contributions to the semiclassical Virasoro blocks  \cite{Fitzpatrick:2016mjq}. The rules governing which classical solutions to gravity should be including in a path integral evaluation are not yet clear, and the boundary description may aid in determining the answer.  It would be illuminating to understand a self-contained prescription for how and when to include different classical solutions for $A_\mu$ with a given set of boundary operator sources when computing the Virasoro blocks \cite{Witten:2010cx}.  Optimistically, understanding the rules in this simpler setting could shed light on the correct procedure for the full correlation function, rather than just the procedure for the individual blocks, potentially identifying subleading gravity configurations associated with the resolution of information loss.  As a first step in this direction, it will be interesting to study the Chern-Simons description of degenerate operators \cite{Witten:2010cx, Gaiotto:2011nm, Fitzpatrick:2016ive} using a version of our formalism with finite-dimensional  sl$(2)$ representations.

\section*{Acknowledgments} 

We would like to thank Hongbin Chen, Ethan Dyer, Guy Gur-Ari, Tom Hartman, Ami Katz, Alex Maloney, Miguel Paulos, Jo\~ao Penedones, Eric Perlmutter,  Steve Shenker, Julian Sonner,  Matt Walters, Huajia Wang, and Jie-qiang Wu for useful discussions, and the GGI for hospitality while parts of this work were completed.  ALF is supported by the US Department of Energy Office of Science under Award Number DE-SC-0010025.  JK  has been supported in part by NSF grants PHY-1316665 and PHY-1454083 and by a Sloan Foundation fellowship.

\appendix

\section{Some Review of Chern-Simons and Holography}
\label{app:ReviewCSandHolography}

In this appendix we will review the Chern-Simons description of gravity, and then we will provide two derivations of the Virasoro Ward identity from Chern-Simons theory.  Some of our analysis of the Virasoro Ward identity follows Verlinde \cite{Verlinde:1989ua}, but we will include details that he left to the reader, and modernize the description in light of AdS/CFT.  For completeness we also explain  how to obtain stress tensor and current correlators directly from the Ward identity.  

\subsection{From Chern-Simons to the Virasoro Algebra}

In this section, we will briefly review how to obtain the Virasoro algebra from the sl$(2)$ Chern-Simons theory.\footnote{An enlarged $W_N$ algebra can be obtained from sl$(N)$ Chern-Simons theory.} The derivation is almost identical to that of a Kac-Moody algebra from the $SU(N)$ CS theory, except that a different boundary condition (equation (\ref{eq:WilsonLineNearBoundary})) is required.  

The (Euclidean) action of the sl$(2)$ Chern-Simons theory is $I_A=I_{\rm CS}+I_{\rm bdy}$, where
\begin{align}\label{Sl(2) CS Action}
I_{\rm CS}[A]
&=\frac{i}{4\pi} \int_Y d^2 x \, d y\,\tilde{\epsilon}^{\mu\nu\lambda}
{\rm Tr} \left(A_\mu \pd_\nu A_\lambda+\frac{2 }{3}A_\mu A_\nu A_\lambda\right)\;,\\
I_{\rm bdy}[A]&=-\frac{1}{8\pi}\int_{\pd Y}d^2 x \sqrt{g}\, \tr\left(A_i A_j g^{ij}\right)\;.\nn
\end{align}
Here the bulk manifold $Y$ has the topology $\mathbb{R}^2\times \mathbb{R}^+=(x^1, x^2)\times y$ with a boundary at $y=0$. We use the Greek letters $\alpha,\beta,\dots$ to denote the bulk coordinates while the Roman letters $i,j,\dots$ denote boundary coordinates, with induced boundary metric $g_{ij}$. 
For future convenience, we also introduce holomorphic coordinates $z=x^1+ i x^2$ and $\bar{z}=x^1- i x^2$. In our convention, the measure in the holomorphic coordinates is $ dz d \bar{z}=2 d x^1 d x^2$\; and 
\be
A_z=\frac{1}{2}\left(A_1-i A_2\right)\;, \quad A_\bz=\frac{1}{2}\left(A_1+i A_2\right)\;.
\ee
We assume that the gauge field $A_\mu$'s are in the fundamental representation $A_\mu=A_{a \mu} t^a$ and the generators
 $t^a$ $(a=\pm, 0)$\footnote{
Notice that in this basis the Killing metric of the sl$(2)$ Lie algebra $\gamma^{ab}={\rm Tr}(t^a t^b)$ is not flat, so one needs to use the Killing metric $\gamma^{ab}$ and its inverse $\gamma_{ab}$ to raise and lower indices.
}
\be\label{SL2 2d Rep}
t^{+}=\left(
\begin{array}{cc}
 0 & 0 \\
 -1 & 0 \\
\end{array}
\right)\;,\quad 
t^0=\left(
\begin{array}{cc}
\frac{1}{2} & 0 \\
 0 & -\frac{1}{2} \\
\end{array}
\right)\;,\quad 
t^{-1}=\left(
\begin{array}{cc}
 0 & 1 \\
 0 & 0 \\
\end{array}
\right)\;,
\ee
satisfy the following commutation relations
\be
[t^+, t^-]=2t^0\;,\quad [t^\pm, t^0]=\pm t^\pm \;.
\ee
Varying the action \eqref{Sl(2) CS Action}, one obtains the EoM: 
\be\label{EoM F}
F_{\mu\nu}\equiv \pd_\mu A_\nu-\pd_\nu A_\mu +[A_\mu, A_\nu]=0\;.
\ee
Noting that the $2d$ boundary metric can always be put into the form of 
\be
g_{ij}(x)=B(x)\eta_{ij}
\ee
the variation of the full bulk plus boundary action is
\begin{align}\label{delta I}
\delta I_A&=\frac{1}{4\pi}\int_{\pd Y}\!{d}^2 x (-\eta^{ij}+i{\tilde \ep}^{ij}){\rm Tr}(A_i \delta A_j)+
\frac{i}{4\pi} \int_Yd^2 x \,d y\,\tilde{\epsilon}^{\mu\nu\lambda}
{\rm Tr} \left(F_{\mu\nu}\delta A_\lambda\right)
\nn\\
&=
-\frac{1}{2\pi}\int_{\pd Y}\!{d}^2 z {\rm Tr}\big(\delta A_z A_{\bar z}\big)\;,
\end{align}
where in the last line we have imposed the on-shell condition \eqref{EoM F},
 and changed to holomorphic coordinates $z=x^1+i x^2$\;.
Furthermore, from \eqref{delta I} we see that the variational principle is well posed once the boundary value for $A_z$ is fixed.

Notice that, assuming that the gauge field $A_\mu$ vanishes at the transverse boundary $|\vec x|\to \infty$ and $y=+\infty$ boundary, the CS action can be rewritten as 
\begin{align}
I_A
&=\frac{i}{4\pi} \int_Yd^2 x \,d y\,\tilde{\epsilon}^{ij}
{\rm Tr} \left(A_y F_{ij}-A_i \pd_y A_j\right)+I_{\rm bdy}[A]\;.
\end{align}
So integrating over $A_y$ is equivalent to imposing the flatness condition $F_{ij}^a=0$\;. This flatness condition leads to $A_i= U^{-1} \pd_i U$, where $U(z,\bz,y)$ is an group element of SL$(2)$. Plugging that into the CS action $I_A$, we find that
\begin{align}\label{sl2 WZW}
I_A \to \Gamma[A_{a z}]
&\equiv\frac{-i }{12\pi} \int_Y d^2 x \,d y\, \tilde{\ep}^{\mu\nu\lambda}\,
{\rm Tr} \bigg[(U^{-1}\pd_\mu U)( U^{-1}\pd_\nu U)(U^{-1}\pd_\lambda U)\bigg]\nn\\
&-\frac{1}{8\pi}\int_{\pd Y}{d}^2 x {\rm Tr}\left(U^{-1}\pd_i U U^{-1}\pd_j U\eta^{ij}\right)\;.
\end{align}
where $A_{a z}(z)t^a=U^{-1}\pd_z U$. This is the chiral sl$(2)$ WZW action! (It is chiral because only $U\to  \Omega(\bz) U$ , $\Omega \in $ SL$(2)$ is a symmetry but $U\to U \Omega(z)^{-1}$ is not.) From the action \eqref{sl2 WZW} the sl$(2)$ Kac-Moody current algebra may be derived \cite{Witten:1983ar}.

In order to get a Virasoro algebra, we should impose a more stringent boundary condition:
\be\label{BC Az}
A_z\Big\vert_{\pd Y} &=&\left( \begin{array}{cc} 
0 &  \frac{T(z)}{k}  \\
-1 & 0 \end{array} \right) \;,
\ee
where $k$ will be identified with the level of the Chern-Simons theory. At this stage we include $k$ in the boundary condition for convenience.
Parameterizing 
\be
A_z &=& b(y)^{-1}\left( \begin{array}{cc} 
0 &  \frac{T(z)}{k}  \\
-1 & 0 \end{array} \right) b(y)\;,\quad
A_{\bar z}
=b(y)^{-1}\left( \begin{array}{cc} 
\frac{1}{2} \omega(z) &  \gamma(z)   \\
-\mu(z) & -\frac{1}{2} \omega(z) \end{array} \right) b(y)\;,
\ee
where $b(y)$ is an arbitrary SL$(2)$ group element vanishing at $y\to \infty$ boundary, it is straightforward to check that the flatness condition $F_{z\bz}=0$ demands that various components of $A_{\bar z}$ be expressed in terms of $T(z)$ via
\begin{align}
-\frac{1}{2}\pd \omega-\frac{1}{k}\mu T+\gamma&=0\;,\\
\pd \mu -\omega&=0\;,\nn\\
\frac{1}{k}{\bar \pd}T-\pd \gamma-\frac{1}{k}\omega T&=0\;.\nn
\end{align}
Following the standard canonical quantization procedure, one can show that $T(z)$ forms a Virasoro algebra with the Lie bracket given by the Dirac bracket \cite{Banados:1998gg}.  This method can  be used to derive  Ward identities via holographic renormalization \cite{deHaro:2000vlm}.

More relevantly for our purpose, we can use the sl$(2)$ Chern-Simons action to compute the correlation functions of the stress tensor.
Consider the path integral with the boundary condition \eqref{BC Az} as the wavefunction
\be
\Phi[T]=\int  [{\cal D} A] e^{-k I_A}
\ee
where $k=\frac{c}{6}$ is the level of the Chern-Simons theory ($c$ is the central charge of the dual CFT), and the measure $[{\cal D} A]$ is understood as 
\be
[{\cal D} A]=\frac{{\cal D}A_y \,{\cal D}A_z \,{\cal D}A_\bz}{\text{Volume of gauge group}}\;.
\ee
The n-point correlation function of stress tensor then is given by 
\be
\langle T(z_1) \cdots T(z_n)\rangle=\int\![{\cal D}T]\big( T(z_1) \cdots T(z_n) \big)\Phi[T]\;.
\ee
Similarly the correlation function of Wilson lines should be understood as 
\be
\langle W[z_1; x_1] \cdots W[z_n; x_n] \rangle=\int\![{\cal D}T]\big( W[z_1; x_1] \cdots W[z_n; x_n]  \big)\Phi[T]\;.
\ee
Using \eqref{delta I} and noting that 
\be
\frac{\delta}{\delta \mu(w)}\left(I_A-\frac{1}{2\pi k}\int\!d^2 z \mu(z) T(z)\right)=-\frac{1}{2\pi k}T(w)\;,
\ee
the n-point correlation function can be rewritten as 
\begin{align}
\langle T(z_1) \cdots T(z_n)\rangle&=\left(-2\pi\right)^n\frac{\delta^n}{\delta \mu(z_1)\cdots \delta \mu(z_n)}\Psi[\mu] \Big\vert_{\mu=0}\;,\\
\Psi[\mu]&=\int\![{\cal D}T]\exp\left(-\frac{1}{2\pi}\int\!d^2 z \mu(z) T(z)\right)\Phi[T]\;.
\end{align}
One immediately recognizes that $\Psi[\mu]$ is Verlinde's geometric Virasoro action \cite{Verlinde:1989ua}, which is related to the sl$(2)$ WZW wave-function $\Phi(T)$  by Legendre transformations \cite{Knizhnik:1988ak}. 
As we will see in the next section, $\Psi[\mu]$ satisfy the source-free Virasoro Ward identity:
\begin{align}\label{V WardIdentity Nosource}
{\cal V}(z) \Psi[\mu]\equiv\Big( \bar{\partial} - \mu(z) \partial - 2 (\partial \mu(z) ) \Big) \left( \frac{\delta}{\delta \mu(z)} \Psi[\mu] \right) + \frac{k}{4\pi} \partial^3 \mu(z) \Psi[\mu]=0\;.
\end{align}

Therefore, in order to obtain the correlation functions of $T$, there is no need to perform the path integral; instead, one can solve \eqref{V WardIdentity Nosource} for the $T$ correlators. For instance, acting $\frac{\delta}{\delta \mu(w)}$ on both sides of \eqref{V WardIdentity Nosource} and then sending $\mu$ to zero, one has 
\be
{\bar \pd}\left( \frac{\delta^2}{\delta \mu(w)\delta \mu(z)} \Psi[\mu] \right)\Big\vert_{\mu=0}=-\frac{k}{4\pi}\pd^3\delta^{(2)}(z-w)\;.
\ee
Then it follows immediately that 
\begin{align}
\Big\langle T(z_1)T(z_2)\Big\rangle&=\left(-2\pi\right)^2\frac{\delta^2}{\delta \mu(z_1)\delta \mu(z_2)}\Psi[\mu] \Big\vert_{\mu=0}
=\frac{3 k}{ (z_1-z_2)^4}\;.
\end{align}
Higher point functions of stress tensor can be obtained in a similar way. 
The Virasoro Ward identity relates correlators with $k+1$ insertion of the stress-energy tensor with $k-$ and $(k-1)-$point functions.   This provides a recursion relation 
\begin{align}
\Big\< T(z) T(z_1) \cdots T(z_k) \Big\> &=  \sum_{i=1}^k \left(\frac{1}{z - z_i} \partial_{z_i} + \frac{2}{(z - z_i)^2}  \right) \Big\< T(z_1) \cdots T(z_k) \Big\>\nn\\
&+\sum_{i=1}^k \frac{c/2}{(z-z_i)^4}
\Big\< T(z_1)\cdots T(z_{i-1})T(z_{i+1})\cdots T(z_k)\Big\>\;.
\end{align}

Before ending this section, we want to comment on the relation between the Chern-Simons theory and the $3d$ gravity.
It is well known \cite{Witten:1988hc} that, 
formulated in terms of vierbeins $e^a\equiv e^{\;\;a}_\mu d x^\mu$ and spin connections $\omega_{a b}\equiv \omega_{\mu a b} d x^\mu$,  the $3d$ Einstein-Hilbert action with a negative cosmological constant
\be
S_{grav}=\frac{1}{16\pi G}\int\! d x^3\sqrt{-g}\left(R+\frac{2}{\ell^2}\right)
\ee
is equivalent to the sl$(2)\times$sl$(2)$ Chern-Simons theory, whose action is given by
\begin{align}
I[A, \bar A]&=I_{CS}[A]-I_{CS}[\bar A]\;,\label{S A Abar}
\end{align}
provided the following identification 
\begin{align}
A_\mu^{\;\;a}&=\frac{1}{\ell}e_\mu^{\;\;a}+\frac{1}{2}\epsilon^{abc}\omega_{\mu b c}\;,
\quad
{\bar A}_\mu^{\;\;a}=-\frac{1}{\ell}e_\mu^{\;\;a}+\frac{1}{2}\epsilon^{abc}\omega_{\mu b c}\;\label{A Abar}\;,
\end{align}
where $\epsilon$ is the Levi-Civita tensor.
Thus the metric is given by 
\be
g_{\mu\nu}=\frac{1}{2}{\rm Tr}\Big((A_\mu-{\bar A}_\mu)(A_\nu-{\bar A}_\nu)\Big)\;.
\ee

The Einstein equation is equivalent to the flatness conditions of Chern-Simons theory $F_{\mu\nu}={\bar F}_{\mu\nu}=0$\;.
Imposing the boundary condition 
\begin{align}\label{BC A barA}
A_z\Big\vert_{\pd Y} &=t^+-\frac{L(z)}{2}t^-\;,\quad A_\bz\Big\vert_{\pd Y}=0\;,\\
{\bar A}_z\Big\vert_{\pd Y}&=0\;,\quad {\bar A}_\bz\Big\vert_{\pd Y} =t^--\frac{{\bar L}(\bz)}{2}t^+\;,
\end{align}
one can parametrize the flat connection $A, {\bar A}$ as 
\begin{align}
A &= \left(v(y)^{-1}\left( \begin{array}{cc} 
0 &  -\frac{L(z)}{2}  \\
-1 & 0 \end{array} \right) v(y)\right)d z\;,\nn\\
{\bar A}
&= \left(v(y)\left( \begin{array}{cc} 
0 & 1  \\
 \frac{{\bar L}(\bz)}{2} & 0 \end{array} \right) v(y)^{-1}\right)d \bz\;,
\end{align}
where $v(y)=e^{- t^0 \log y}$\;.
The corresponding metric takes the form of \eqref{AAdS3 metric}. This is the most general solution of the 3d Einstein Equation which has asymptotic AdS$_3$ geometry. Of course one can also consider more general boundary conditions than \eqref{BC A barA}. That is, making $A_\bz, {\bar A}_z$ non-vanishing while keeping $A_z$ and ${\bar A}_\bz$ unchanged; as we showed before, such boundary conditions will lead to two copies of Virasoro algebra, with $T(z)=-\frac{c}{12}L(z)$ (and ${\bar T}(\bz)=-\frac{c}{12}{\bar L}(\bz)$) being the holomorphic (and anti-holomorphic) stress tensor of the boundary dual CFT. 

Now consider some matter field $\phi$ in the bulk gravitational theory with conformal weights $(h, \bar h)$. If one wants to represent this matter field by Wilson lines, the Wilson lines must include both $A$ and $\bar A$ gauge field. However, by construction $\bar A$ only depends on the antiholomorphic coordinates $\bz$. Thus if we are only interested in  the holomorphic sector of the correlators, it suffices to keep only $A$ in Wilson lines and in the action.  This justifies our method to compute the CFT correlators (holomorphic part) from one copy of Chern-Simons  theory.

\subsection{Review of the Virasoro Ward Identity}

Let us begin by briefly reviewing the usual statement of the Virasoro Ward identity for correlators as it arises from CFT.  We will then reformulate it as a statement about the generating function of correlators, since that form is more natural in the Chern-Simons description.  For a more detailed treatment see e.g. \cite{DiFrancesco:1997nk, Ginsparg}.

Recall that the current associated with conformal transformations $x^\mu \rightarrow x^\mu + \epsilon^\mu$ is
\be
J^\mu &=& T^{\mu\nu} \epsilon_\nu
\ee
where $\epsilon_\nu$ is a conformal transformation, i.e. it satisfies
\be
(\partial_\mu \epsilon_\nu + \partial_\nu \epsilon_\mu) = \frac{2}{d} (\partial \cdot \epsilon) \eta_{\mu\nu} .
\ee
In $d=2$, this implies that $\epsilon^z$ is holomorphic, $\bar{\partial} \epsilon^z=0$, and $\epsilon^{\bar{z}}$ is anti-holomorphic.
The Ward identity follows from the fact that $\bar{\partial} J$ vanishes up to contact terms, which act on local operators to generate conformal transformations:
\be
\frac{1}{2\pi} \int d^2z \< \epsilon(z) \bar{\partial} T(z) \Phi_1(y_1) \dots \Phi_n(y_n)\> &=& \delta_{\epsilon} \< \Phi_1(y_1) \dots \Phi_n(y_n)\>
\ee
The infinitesimal conformal transformation of a Virasoro primary operator is
\be
\delta_\epsilon \Phi(w)  &=& (\epsilon(w) \partial + h_\Phi \partial \epsilon(w) )\Phi(w),
\ee
whereas the infinitesimal conformal transformation of the stress tensor $T(z)$ is
\be
\delta_\epsilon T &=& \epsilon \partial T + 2 ( \partial \epsilon) T + \frac{c}{12} \partial^3 \epsilon .
\ee
So for any correlation function of primary operators and stress tensors we have 
\be
\frac{1}{2\pi} \int d^2z \< \epsilon(z) \bar{\partial} T(z) ( \dots ) \> &=& \frac{1}{2\pi} \int d^2z \left\< \Big(\epsilon \partial T(z) + 2 (\partial \epsilon) T(z) + \frac{c}{12} \partial^3 \epsilon\Big) \frac{\delta}{\delta T(z)} ( \dots )\right\> \nn\\
 &+& \sum_i \left\< \Big(\epsilon(z_i) \partial_{z_i}   + h_i \partial \epsilon(z_i) \Big) (\dots)\right\>
\label{eq:UsualWardIdentity}
\ee
where the sum on $i$ is over the primary operators.
This Ward identity can be re-written as a statement about 
\be
\Psi[\mu; z_i] &=& \< \Phi_1(z_1) \dots \Phi_N(z_N) e^{i \int d^2 z \mu(z, \bar{z})) T(z)} \>
\ee
which is the generating functional of the correlator of some specific set of Virasoro primaries and any number of stress tensors.
It takes the form
\be
&& \left( \bar{\partial} - \mu(z) \partial - 2 (\partial \mu(z) ) \right) \left( \frac{\delta}{\delta \mu(z)} \Psi[\mu; z_i] \right) + \frac{c}{12}  \frac{\partial^3 \mu(z)}{2 \pi i} \Psi[\mu; z_i]
\nn\\
&& =  \sum_i \left( h_i \partial \delta^2(z-z_i) + \delta^2(z-z_i) \partial_{z_i} \right) \Psi[\mu; z_i] 
\ee
where the explicit derivatives $\partial$ and $\bar \partial$ all act on $z$.  The relationship with the prior version of the Ward identity is easy to see if we note that $\mu =  -i \frac{\delta}{\delta T}$ and $T = -i \frac{\delta}{\delta \mu}$.   In both AdS/CFT and the Chern-Simons description, we view $\mu$ as a boundary source for $T$; it will either appear as a deformation $\mu \, d \bar z^2$ of the boundary metric or a component of the C-S field $A_{\bar z}$.

\subsection{The Virasoro Ward Identity from Chern-Simons Theory}
\label{app:VirasoroWardfromCS}

In this subsection, we will discuss how to derive the Virasoro Ward identity from an sl$(2)$ Chern-Simons theory.  
In the modern AdS/CFT language, the boundary Ward identity follows from a study of bulk gauge transformations (diffeomorphisms) that do not vanish near the boundary.  In the presence of operators inserted on the boundary, the asymptotic symmetry relations become the ward identities for the boundary dual theory, see e.~g.~\cite{deHaro:2000vlm} for detailed discussions. The Virasoro Ward identity we are interested in here is an example of the above idea in the context of AdS$_3$/CFT$_2$

However, in the remainder of this subsection we will discuss an older derivation:  following Verlinde \cite{Verlinde:1989ua}, one can view the Virasoro Ward identity as the gauge-independence constraint on the generating function of correlation functions (which can also be viewed as a wavefunctional).

For starters, consider the sl$(2)$ Chern-Simons action \eqref{Sl(2) CS Action} on the bulk manifold $Y$.
Verlinde's analysis \cite{Verlinde:1989ua} is based on canonical quantization, where we interpret the $y$-direction as time. 
Since $y$ plays the role of time, the field $A_y$ is not  dynamical, as it does not have a conjugate momentum.  Thus $A_y$ is simply a Lagrange multiplier implementing the constraint that $F_{z \bar z}$ vanishes in the absence of sources. 
 The other fields $A_z$ and $A_{\bar z}$ are dynamical, and are canonically conjugate variables. However, there are still gauge redundancies, and we have to eliminate the extra degrees of freedom by `moding out' by gauge transformations.

First we parametrize the gauge field as 
\be
A=b(y)^{-1}(a_z dz+a_\bz d\bz) b(y)\;,
\ee
where $b(y)$ is an arbitrary SL$(2)$ group element vanishing at $y\to \infty$ boundary, and $a_z$ and $a_\bz$ are now representing boundary degrees of freedom living on $y=0$ surface. That accounts for `moding out'  the bulk (pure) gauge transformations by completely fixing the $y-$dependence.
It is convenient to use the variables
\begin{align}\label{az boundary condition}
a &= \left( \begin{array}{cc} 
\frac{1}{2} \omega_z & e^-_{z}  \\
 -e^+_{z} & -\frac{1}{2} \omega_z \end{array} \right)  dz
+\left( \begin{array}{cc} 
\frac{1}{2} \omega_{\bz} & e^-_{\bz}  \\
 -e^+_{\bz} & -\frac{1}{2} \omega_\bz \end{array} \right)  d \bz
\end{align}
 In these variables the canonical commutation relations following from the form of classical action are
\begin{align}
\left[ \omega_z(z), \omega_{\bar w}(w) \right] &= \frac{4 \pi}{k} \delta^{(2)}(z-w)\;,
\nn \\
\left[ e^-_z(z), e^+_{\bar w}(w) \right] &=-\frac{2 \pi}{k} \delta^{(2)}(z-w)\;,
\nn \\
\left[ e^+_z(z), e^-_{\bar w}(w) \right] &= -\frac{2 \pi}{k} \delta^{(2)}(z-w)\;.
\end{align}
Now we can compute the wavefunctional of the sl$(2)$ Chern Simons theory at $y=0$. It is just a path integral of the Chern-Simons action subject to the boundary condition of equation \eqref{az boundary condition}:
\begin{align}\label{def Psi}
\Psi[e^+_z, e^+_{\bar z}, \omega_z; z_i, x_i]=
\int \! [{\cal D} A] \big(\Pi_i  e^{-\int A_{(i)}}\big)e^{-k I_A}\;.
\end{align}
Notice that the wavefunctional cannot depend on all $a_z$ and $a_\bz$ components, just like the wave-function in quantum mechanics cannot be simultaneously a function of positions and momenta. Here for calculational convenience we have chosen as our canonical coordinates $e^+_z, e^+_{\bar z}, \omega_z$ (referred to as the `mixed polarization'); the other field variables not appearing in $\Psi$ are their conjugate momenta.  In this basis, they act as differential operators 
\be\label{Momenta Variable}
\omega_{\bar z}(z) = \frac{4 \pi}{k} \frac{\delta}{\delta \omega_z(z)}, \ \ \ e^-_z =-\frac{2 \pi }{k} \frac{\delta}{\delta e_{\bar z}^+ (z)}, \ \ \ e^-_{\bar z}=- \frac{2 \pi }{k} \frac{\delta}{\delta e_{z}^+ (z)}\;.
\ee
In the expression for $\Psi$ we have also included a product of Wilson lines emanating from the boundary, where the gauge field of the $i^{\mathrm{th}}$ Wilson line is in the representation $x_i$. Thus the wavefunctional $\Psi$ also depends on the locations $z_i$ and the representations (labeled by $x_i$) of the Wilson lines, as pictured in figure \ref{fig:SomeWilsonLinesNearTheBoundary}.

We also need to mod out by boundary gauge transformations. This can be done by demanding the wavefunctional $\Psi$ satisfy some constraint equations. This is similar to the case of QED, where one defines physical states by demanding they obey a Gauss's law constraint.  In the absence of Wilson lines, the equations of motion set 
\be
{\cal F}_{z \bz} =\pd a_\bz-{\bar \pd}a_z+[a_z, a_\bz]= 0
\ee
 in Chern-Simons theory. 
At the quantum level, this is reflected by the fact that the field strengths ${\cal F}_{z \bar z}^a$ are the generators of infinitesimal gauge transformations (see e.g. \cite{Carlip:2005zn} for a relevant review).  This means that  the operator ${\cal F}_{z \bar z}^a$ must annihilate the wavefunctional of the sl$(2)$ theory \cite{Labastida:1989wt}. 

In the presence of Wilson lines the constraint equations that the wavefunctional $\Psi$ obey should be modified accordingly: 
\be
F_{z \bar z}^- \Psi  &=& \frac{k}{2\pi} \left(-\partial_z e_{\bar z}^+ + \partial_{\bar z} e_z^+ + \omega_z e_{\bar z}^+-\frac{4\pi}{k} e_z^+ \frac{\delta}{\delta \omega_z}\right)\Psi 
= -\sum_i \delta^{(2)}(z-z_i) L^-_i\Psi\;,
\nn \\
F_{z \bar z}^{+}\Psi  &=& \left(\partial_z \frac{\delta}{\delta e^+_z} - \partial_{\bar z} \frac{\delta}{\delta e^+_{\bar z}}
+ \omega_z \frac{\delta}{\delta e^+_{z}} - \frac{4 \pi }{k} \frac{\delta}{\delta \omega_{z}} \frac{\delta}{\delta e^+_{\bar z}}
\right)\Psi
= -\sum_i \delta^{(2)}(z-z_i) L^{+}_i \Psi\;,
\nn \\
F_{z \bar z}^{0}\Psi  &=&\left(-\frac{k}{4 \pi} \partial_{\bar z} \omega_z + \partial_z \frac{\delta}{\delta \omega_{z}} - e^+_z \frac{\delta}{\delta e^+_{z}} + e^+_{\bar z} \frac{\delta}{\delta e^+_{\bar z}} 
\right)\Psi
= -\sum_i \delta^{(2)}(z-z_i) L^{0}_i\Psi 
\ee
Here the $L^a_i$ is the sl$(2)$ generator in the $x_i$  representation acting on the $i^{\mathrm{th}}$ Wilson line, which ends at $z_i$ on the boundary. We have chosen our canonical coordinates and replaced the canonical momentum variables in ${\cal F}_{z \bz}^a$ using the differential operators in equation \eqref{Momenta Variable}.

One can use the $F_{z\bz}^-$ and $F_{z\bz}^0$ constraint equation to solve algebraically for $\frac{\delta }{\delta \omega_z} \Psi $ and  $\frac{\delta }{\delta e_z^+} \Psi $. Furthermore, one should impose the boundary condition \eqref{BC Az} --- as we argued before, this is all we need to obtain the Virasoro algebra.  After plugging in the expression of $\frac{\delta }{\delta \omega_z} \Psi $ and  $\frac{\delta }{\delta e_z^+} \Psi $ and setting $e_z^+=1$ and $\omega_z=0$, the ${\cal F}_{z\bz}^+$ constraint equations on the wavefunctional $\Psi$ becomes: 
\begin{align}\label{Virasoro Identity with source}
&\Big( \bar{\partial} - \mu(z) \partial - 2 (\partial \mu(z) ) \Big) \left( \frac{\delta}{\delta \mu(z)} \Psi \right) + \frac{k}{4\pi} \partial^3 \mu(z) \Psi \nn\\
&-\left(\frac{1}{2}\pd^2 \delta_i L^-
+\pd \delta_i L^0
+\delta_i L^+
\right)\Psi
+\frac{2\pi}{k}(\delta_i L^-)\left( \frac{\delta}{\delta \mu(z)} \Psi \right)=0\;,
\end{align}
where we have denoted $e^+_\bz\equiv \mu$ to be consistent with the notation in the previous section, and $\delta_i L^a$ is shorthand for $\delta_i L^a\equiv \sum_i \delta^{(2)}(z-z_i)L_i^a$\;. We recognize that the first line of \eqref{Virasoro Identity with source} agree with the sourceless Virasoro Ward identity \eqref{V WardIdentity Nosource}\;. And the localized source term in the second line stems from the inclusion of Wilson lines in the path integral. The Virasoro Ward identity \eqref{Virasoro Identity with source} we derived from sl$(2)$ Chern-Simons theory agrees with Verlinde's Equation (4.8)  in \cite{Verlinde:1989ua}, except for the very last term.

Now let us establish a very important fact about the relationship between the physical coordinates $z_i$ and the `internal' coordinates $x_i$, which have been introduced purely to encode information about the infinite dimensional sl$(2)$ representation.  A priori, there is no connection between the $x_i$ and $z_i$.  However, from the definition of the wavefunction $\Psi$, Equation \eqref{def Psi}, one sees that 
\begin{align}
\partial_{z_i}\Psi[\mu; z_i, x_i] \Big \vert_{x_i=0}= -\left(L_i^++T(z_i)L_i^-\right)\Psi[\mu; z_i, x_i]\Big\vert_{x_i=0} = -\pd_{x_i}\Psi[\mu; z_i, x_i] \Big \vert_{x_i=0}\;. 
\label{eq:SpacetimeandInteralCoordsIdentified}
\end{align}
In the first equality we have used the fact that the gauge field in the $i^{\mathrm{th}}$ Wilson line is parametrized as $A=L_i^++T(z_i)L_i^-$, where the sl$(2)$ generators of the infinite representation are given by 
\be
L^+_i = \partial_{x_i}\;,\quad L_i^0 = x_i \partial_{x_i} + h_i\;,\quad  L^{-}_i = \frac{1}{2} x_i^2 \partial_{x_i} + h_i x_i\;;
\ee
and in the second equality we have used $L^-_i \to 0$ as $x_i \to 0$.
Therefore, in the vicinity of $x_i = 0$  we can simply replace derivatives on $x_i$ with derivatives with respect to $z_i$.  This indicates that $\Psi$ represents a correlator of primary operators of dimension $h_i$ at $z_i$, with the $L_i^a$ acting as the global Virasoro generators on $z_i$.

Setting $x_i=0$, the equation \eqref{Virasoro Identity with source} reduces to 
\begin{align}
&\Big( \bar{\partial} - \mu(z) \partial - 2 (\partial \mu(z) ) \Big) \left( \frac{\delta}{\delta \mu(z)} \Psi \right) + \frac{k}{4\pi} \partial^3 \mu(z) \Psi \nn\\
&=\sum_i\Big(
h_i\pd \delta^{(2)}(z-z_i) 
-\delta^{(2)}(z-z_i)\pd_{z_i}
\Big)\Psi
\;.
\end{align}
This is just the Virasoro Ward identity stated in the form of equation (\ref{eq:StatementVerlindeTypeWardIdentity}).
In order to get more intuition, let us work out the Virasoro Ward identity with one insertion of stress tensor. Setting $\mu = 0$ and noting that $\frac{\delta }{\delta \mu(z)} \to -\frac{1}{2\pi} T(z)$, one has 
\be
 -\frac{1}{2\pi} \bar \partial_z \big(T(z) \Psi\big)
= 
-\frac{1}{2\pi}{\bar \pd}\sum_i\Big(-
h_i\pd \frac{1}{z-z_i}
+
\frac{1}{z-z_i}\pd_{z_i}
\Big)\Psi\;.
\label{eq:ZoomedWI}
\ee
Now stripping of $\bar \pd$ and the overall factor on both side, and interpreting $ T \Psi$ as the  $\< T(z) X \>$ correlator where $X$ is a product of local primaries, then the above equation becomes the familiar Ward identity
\be
\label{eq:VirasoroRecursionRelation}
\big\<T(z) X\big\>
= 
\sum_i\Big(-
h_i\pd \frac{1}{z-z_i}
+
\frac{1}{z-z_i}\pd_{z_i}
\Big)\<X\>\;.
\ee

\section{Details of Perturbation Theory and the Shadow Formalism}
\label{app:PerturbationTheoryandShadows}

In this appendix we collect some technical details on  the relation between different gauge choices for the Chern-Simons propagator and the shadow formalism in CFT$_2$.

\subsection{Connection with Covariant Gauges}
\label{app:ConnectionwithCovariantGauge}

Chern-Simons perturbation theory has been studied in covariant gauges \cite{Guadagnini:1989am, Guadagnini:1989kr}.  These works were primarily motivated by knot theory, and so they aimed at computing closed Wilson loops in a full three dimensional Euclidean space, rather than open Wilson lines that end on boundary surfaces.  Thus it was natural for them to use covariant gauge fixing terms and introduce ghosts via the usual Fadeev-Popov procedure;  they found \cite{Guadagnini:1989am} only finite quantum corrections.  For completeness, let us explain the connection between their gauge choice and our prescription, which has been motivated by AdS/CFT.

In Lorentz gauge $\nabla^\mu A_\mu^a = 0$, the Chern-Simons propagator takes the form \cite{Guadagnini:1989am}
\be
\left\<A_\mu^a(x) A_\nu^b(0) \right\>_L =  \frac{i \delta^{ab}}{k} \frac{\epsilon_{\mu \nu \rho} x^\rho}{|x|^3}
\ee
Thus in Lorentz gauge $\< A_z A_z \>_L = 0$ identically.
We can relate this form of the propagator to ours by performing a gauge transformation $A_\mu \to A_\mu + \partial_\mu \phi$ in order to set $A_y = 0$.  For this purpose, we must choose $\phi$ as
\be
\phi(z, \bar z, y) = -\int_{\infty}^y A_y(z, \bar z, y') dy'
\ee
assuming that in the original Lorentz gauge, $A_y \to 0$ at infinity, as is consistent with its 2-pt correlator.  After this gauge transformation we find
\be
\< A_z(x_1) A_z(x_2) \> &=&  \int^{y_2} dy' \partial_{z_2} \left\< A_z(x_1) A_y(z_2, y') \right\>_L   - \int^{y_1} dy'  \partial_{z_1} \left\< A_y(z_1, y') A_z(x_2) \right\>_L 
\nn \\
&=& \frac{i \delta^{ab}}{k} \left( \int^{y_2} dy' \partial_{z_2} \frac{\bar z_{12}}{\left( z_{12} \bar z_{12} + (y_1 - y')^2 \right)^{3/2} }
- \int^{y_1} dy' \partial_{z_1} \frac{\bar z_{12}}{\left( z_{12} \bar z_{12} + (y_2 - y')^2 \right)^{3/2}} \right)
\nn \\
&=&  \frac{2 i \delta^{ab}}{k } \frac{1}{(z_1 - z_2)^2} 
\ee
as expected;  all $\bar z$ and $y$ dependence has cancelled.  A similar calculation shows that $\<A_z A_{\bar z} \> = 0$ after the gauge transformation.  This is how our gauge field propagator can be recovered from Lorentz gauge.  Notice that in this gauge, integrals over $z$ along a Wilson line can develop UV divergences from the region $z_{12} \to 0$, though there were no such singularities in the covariant gauge \cite{Guadagnini:1989am, Guadagnini:1989kr}.

\subsection{Shadow Correlators as Generating Functions for the OPE}
\label{app:ShadowDetails}

In this section we will show very explicitly how the shadow formalism acts a projector onto a single representation of the sl$(2)$ global conformal group.  Let us consider the correlator of a shadow operator $\tilde \CO_5$ with two primaries:
\be
\< \CO_1(z_1) \CO_2(z_2) \tilde \CO_5(z_5 + x_5) \> = \frac{1}{z_{12}^{h_1+h_2+h_5-1} (z_1 - z_5 - x_5)^{h_1 - h_2 - h_5 +1} (z_2 - z_5 - x_5)^{h_2 - h_1 - h_5 + 1} }
\nn
\ee
This is a kind of generating function for the OPE.  To see this, note that the OPE  is
\be
\CO_1(z) \CO_2(0) = \frac{1}{z^{h_1+h_2-h_5}} \sum_{k=0}^\infty a_k z^{k}  \partial^k \CO_5(0) 
\ee
If we take the correlator of this with $\tilde \CO_5(x)$ we obtain
\be
\< \CO_1(z) \CO_2(0) \tilde \CO_5(x) \> = \frac{1}{z^{h_1+h_2-h_5}} \sum_{k=0}^\infty a_k z^{k}  \delta^{(k)}(x)
\ee
In this sense, the 3-pt function with a shadow operator is a generating function for the (global) OPE coefficients.  But to further clarify the situation, we will derive the global or sl$(2)$ conformal block decomposition from the shadow formalism.  Note that
\be
\< \CO_5(X) \CO_1(z) \CO_2(0) \> = \frac{1}{z^{h_1+h_2-h_5}} \sum_{k=0}^\infty a_k z^{k}  \partial^k \< \CO_5(X) \CO_5(0)  \>
\ee
This means that
\be
\frac{1}{(X-z)^{h_5 + h_1 - h_2} X^{h_5 + h_2 - h_1}} = \sum_{k=0}^\infty a_k z^{k}  \partial^k \frac{1}{X^{2h_5}}
\ee
We can differentiate $n$ times with respect to $z$ and set $z \to 0$ and $X \to \infty$ to find
\be
a_n = \frac{1}{(2h_5)_n n!} (h_5 + h_1 - h_2)_n
\ee
Thus we see that the $a_n$ are OPE coefficients divided by the normalization factors $\< h_5 | L_1^n L_{-1}^n | h_5 \>$.  When we integrate against $\< \CO_3 \CO_4 \CO_5 \>$ to extract a conformal block, we obtain these normalizations and one set of OPE from the shadow operator $\tilde \CO_5$, while the $\CO_5$ terms provide the other OPE coefficients.  Explicitly, we obtain the global conformal block from
\be
G & = & \int dx \< \CO_1(\infty) \CO_2(1) \CO_5(x) \> \<  \tilde  \CO_5(x) \CO_3(z) \CO_4(0)  \>
\nn \\
& = & \int dx  \< \CO_1(\infty) \CO_2(1) \CO_5(x) \>  \frac{1}{z^{h_1+h_2-h_5}} \sum_{k=0}^\infty a_k z^{k}  \delta^{(k)}(x)
\nn \\
& = &  \int dx  \delta(x) \sum_{k=0}^\infty  (-\partial)^k \< \CO_1(\infty) \CO_2(1) \CO_5(x) \>  \frac{\< h_5 | L_1^k \CO_3(z) \CO_4(0) \>} {\< h_5 | L_1^k  L_{-1}^k | h_5 \>}
\nn \\
& = &  \sum_k  \< \CO_1(\infty) \CO_2(1) L_{-1}^k | h_5 \>  \frac{\< h_5 | L_1^k \CO_3(z) \CO_4(0) \>} {\< h_5 | L_1^k  L_{-1}^k | h_5 \>}
\ee
and this last line is the definition of a holomorphic global conformal block in CFT$_2$.  Note that if we restrict the domain of $x$ integration to the region $x \in [0,z]$ we (formally) obtain the same result.

\section{Regulating Divergences}
\label{app:Regulation}

In this appendix we discuss the important and thorny question of regulating divergences from singular terms in the $T(z_i) T(z_j)$ OPE.  We preserve a proposal from Version 1 of this paper in section \ref{app:FreeBoson}, which was sufficient for the computations in the body of the paper.  Then in section \ref{app:FullRegulator}, new to Version 2 of this paper, we discuss a better regulator that we believe works to all orders, and (in a certain sense) has passed checks up to order $\frac{1}{c^5}$.  As we will explain, this regulator still has some unsatisfying features from both a conceptual and  a computational perspective, and there is room for further improvements.

\subsection{(Naive) Free Boson Regulator}
\label{app:FreeBoson}

We would like to define a convenient regulator for stress tensor OPE singularities $T(z_i) T(z_j)$ to eliminate divergences from integrals over $z_i, z_j$ on the same Wilson line.  A natural approach would be to normal order the stress tensors.  However, if we express $T(z) = \sum_n z^{2+n} L_n$ in terms of $L_n$, it is unclear how to define a normal ordering procedure for a product like $L_n L_m$, since the $L_n$ do not commute.  We will now explain a way to obtain a consistent procedure, and then we will show that it has a simple and universal definition.

The multi-stress tensor correlators are determined entirely by the Virasoro Ward identity, so in particular, they are theory-independent.  This means that we can compute their general form by working with any specific theory.  Thus let us consider $N$ copies of a free boson, and construct the stress tensor
\be
T(z) =  \frac{1}{2} \sum_{i=1}^N \partial \phi_i(z) \partial \phi_i(z) .
\ee
This theory has central charge $N$, which we will take to be a parameter.  We define the normal ordered product of stress tensors as normal ordering of the underlying $\phi_i$ bosons.  With this definition, it is trivial to compute correlators like 
\be
\< :T(\infty) T(1): \ :T(z) T(0): \> = \frac{c^2}{4 (1-z)^4}+ \frac{c^2}{4}+ \frac{c}{(1-z)^2} .
\ee
The rule for computing such correlators is to write each $T(z_i)$ in terms of underlying bosons, and drop all terms where bosons are contracted with other bosons inside a single normal ordering symbol. 

In fact, we can define these normal ordered correlators without any explicit mention of the underlying free boson theory.   We just need to characterize the pole structure of the regulated correlator of stress tensors.  We can equivalently characterize the difference between the full correlator and the regulated one:
\be
\delta G(z_i) \equiv \< T(z_4) T(z_3) T(z_2) T(z_1) \>-\< :T(z_4) T(z_3): \ :T(z_2) T(z_1): \>,
\ee
i.e $\delta G(z_i)$ is the terms that the regulator removes. It is uniquely fixed by  demanding that it removes the singularities associated with the OPE of two $T$s from  the same Wilson line, but without ruining the OPE of $T$s from different lines.  We have reintroduced the dependence on all positions because this is necessary to make the pole structure visible in all channels.  The subtraction  $\delta G(z)$ is fixed by demanding that it removes all of the $z_{12}$ and $z_{34}$ singularities, but does not introduce  $z_{ij}^{-3}$ (or worse) singularities in the other channels.  To see this explicitly, write down the most general form of $\delta G$ allowed by scaling together with the fact that all terms must be singular at $z_{12} \rightarrow 0$ and $z_{34} \rightarrow 0$:
\be
\delta G(z_i) &=& \frac{c^2}{4 z_{12}^4 z_{34}^4} + \frac{c}{z_{12}^2 z_{34}^2} \left( \frac{A}{z_{23}^2 z_{24}^2 } + \frac{B}{z_{23}^2 z_{14}^2 } + \frac{C}{z_{13}^2 z_{24}^2 }+ \frac{D}{z_{13}^2 z_{14}^2 } \right).
\ee
No explicit $z_{12}^{-1}$ or $z_{34}^{-1}$ terms are allowed since then scaling would require a $z_{ij}^{-3}$ term to compensate.  However, in a series expansion around $z_{12}\sim 0$, there {\it are} $z_{12}^{-1}$ terms, and similar there are $z_{34}^{-1}$ terms in a $z_{34} \sim 0$ expansion. Demanding that these singular terms exactly match those in the full correlator $\< T TTT\>$ fixes the coefficients $A,B,C,D$ uniquely to give
\be
\delta G(z_i) &=& \frac{c^2}{4 z_{12}^4 z_{34}^4} + \frac{c}{z_{12}^2 z_{34}^2} \left( \frac{1}{ z_{23}^2 z_{14}^2} + \frac{1}{z_{13}^2z_{24}^2} \right),
\ee
which reproduces the free boson regulator. 

\subsection{Regulation to All Orders}
\label{app:FullRegulator}

As we emphasized in section \ref{sec:VirasoroWardIdentity}, perhaps the most crucial test of any regulation scheme is that it preserves the Virasoro Ward identity.  We reviewed a proof of this identity in appendix \ref{app:ReviewCSandHolography}, but that argument is merely formal as it ignores divergences.  Instead of going through the proof and looking for subtleties, we can test the Ward identity in a much more direct and practical way by computing the correlators of Wilson lines with any number of stress tensors.  A valid regulator must lead to the identity
\be
\label{eq:PracticalWardIdentity}
\< T(z_1) \cdots T(z_n) W_h(z,0) \> = \< T(z_1) \cdots T(z_n) \CO_h(z) \CO_h(0) \>  
\ee
where $\CO_h$ is a Virasoro primary operator with holomorphic dimension $h$, and we include any number $n$ of stress tensors.  If this identity holds, then $W_h$ has the correct matrix elements with all products of Virasoro generators, and thus it must correctly reproduce the vacuum OPE block.  Then our prescription for using Wilson lines to compute general OPE blocks should also be exact.

The right hand side of equation (\ref{eq:PracticalWardIdentity}) is uniquely defined by (and can be conveniently computed from) the recursion relation of equation (\ref{eq:VirasoroRecursionRelation}).  The left-hand side is computed by expanding the exponential defining $W_h(z,0)$, which leads to integrals over multi-stress tensor correlators of the form
\be \label{eq:RegulatedStressTensorCorrelators}
\< T(z_1) \cdots T(z_n) [T(y_1) \cdots T(y_m)] \>
\ee
where we use $[\cdots]$ to denote regulated products of stress tensors.  Thus at an operational level, we need to provide a definition for the correlators of equation (\ref{eq:RegulatedStressTensorCorrelators}) that eliminates OPE singularities when $y_i \to y_j$, as these singularities produce divergences when we integrate over the $y_i$.

The free boson regulator of appendix \ref{app:FreeBoson} was a definition for equation (\ref{eq:RegulatedStressTensorCorrelators}) that reproduced equation (\ref{eq:PracticalWardIdentity}) to leading and sub-leading order in $1/c$, but it fails at higher orders.  We first noticed the problem by studying the exact Virasoro OPE block to the first few orders in $z$:
\be
\left[ \CO_h(z) \CO_h(0)\right]_{\rm vac.block} |0\> &=& z^{-2h} \left( 1 + \frac{2h}{c} z^2 L_{-2} + \frac{h}{c} z^3 L_{-3} \right. \\
 && \left. + z^4\left( \frac{2h(1+5h)}{c(22+5c)} L_{-2}^2 + \frac{3 (4+c-2h)h}{c(22+5c)} L_{-4} \right)+ \dots \right) | 0 \>  \nn
 \label{eq:OPEseries}
\ee
Note that at order $z^4$, we have non-trivial denominators with poles at $c = -\frac{22}{5}$ due to null descendants of the identity.  The free boson regulator does not reproduce this pole structure correctly.  Ultimately, this is due to the fact that this regulator propagates extra states in the free boson Hilbert space that do not correspond to Virasoro modes.

In section \ref{app:TheRegulator} we will define a regulator that provides a prescription for equation (\ref{eq:RegulatedStressTensorCorrelators}) that appears to reproduce equation (\ref{eq:PracticalWardIdentity}) exactly to order $z^{10}$, which means that it passes checks including up to $5$ stress tensors, as explained in section \ref{app:OrderbyOrderTest}.  We also verify the regulator to all orders in $z$ with two stress tensors in section \ref{app:Testwith2Ts}.  However, we do not have an all-orders proof.

Aside from the fact that we have not proven that it works, our regulator has some potentially unsatisfying features.  One is simply that it is an ad hoc prescription, rather than something systematic based on adding specific counter-terms to an action according to a familiar recipe.  It would be very interesting to pursue such an approach in the future.

Another issue with our regulator is that it leads to $\<W_h(z,0)\> = \< \CO_h(z) \CO_h(0) \>$ exactly.  For the infinite dimensional representations of sl$(2)$ that we are studying here, this is not a problem.  But in a very interesting recent paper \cite{Besken:2017fsj}, the Wilson line formalism was used to attempt to derive the dimensions of degenerate states as gravitational self-energies.  From this point of view, the finite dimensional representations of sl$(2)$ should inherit the $c$-dependent dimensions of the degenerate states from a computation of $\<W_h(z,0)\>$ in $1/c$ perturbation theory.   But our regulator will automatically set $\< [T(y_1) \cdots T(y_n) ] \> = 0$, so that the gravitational self-energies vanish.  Thus our regulator does not shed much light on the issues encountered in \cite{Besken:2017fsj}.

A final unsatisfying feature of our regulator is that it does not make the computation of higher order $\frac{1}{c^n}$ corrections to Virasoro blocks particularly straightforward.  The problem is that while we will give a very simple presciption for all correlators of the form of equation (\ref{eq:RegulatedStressTensorCorrelators}), we do not have a simple way of computing correlators like $\< [T(x_1) \cdots T(x_n)]  [T(y_1) \cdots T(y_m)] \>$ that involve regulated stress tensors on both sides, or correlators of $[T(y_1) \cdots T(y_m)]$ with other local Virasoro primaries.  Both types of correlators are implicitly determined by  equation (\ref{eq:RegulatedStressTensorCorrelators}), but they are not easy to compute beyond the leading order in $1/c$ (to leading order they are just disconnected stress tensor correlators).  

Our primary goal in this paper was to take the sl$(2)$ Wilson line formalism and find a concrete prescription for the exact Virasoro OPE blocks (and thus of Virasoro conformal blocks more generally), and it appears that our regulator is sufficient for this purpose.  We have  seen how to do several explicit computations, and we will discuss a general systematic algorithm below.  But it would be much preferable to have a less implicit definition to streamline higher-order computations, as calculational benefits usually correlate with conceptual advances.

\subsubsection{A Proposal for the Regulator}
\label{app:TheRegulator}

Our regulator has a simple definition directly in terms of equation (\ref{eq:RegulatedStressTensorCorrelators}).  We choose
\be 
\< T(z_1) \cdots T(z_n) [T(y_1) \cdots T(y_m)] \> \overset{n<m}{=} 0,
\ee
or in words,  the correlator vanishes when $n < m$.  Then when $n\geq m$ we define
\begin{equation}
\< T(z_1) \cdots T(z_n) [T(y_1) \cdots T(y_m)] \> = \sum_{\textrm{groups } (z_{i_{j,1}}, \dots, z_{i_{j,s_j}} ,y_j)} \prod_{j=1}^m \< T(z_{i_{j,1}}) \dots T(z_{i_{j,s_j}}) T(y_j)\> ,
\label{eq:RegProp}
\end{equation}
where by $(z_{i_{j,1}}, \dots, z_{i_{j,s_j}} ,y_j)$ we are indicating a sum over groupings.
In words, this means that we sum over all groupings of the $T$s into $m$ groups each containing exactly one $T(y_i)$, and the contribution of each group is simply the standard multi-point correlator of stress tensors.  This means that in the special case $n = m$, equation (\ref{eq:RegulatedStressTensorCorrelators}) will simply be a sum of products of  2-pt correlators between $\< T(z_i) T(y_j) \>$.  For example
\be 
\< T(z_1) T(z_2) T(z_3) [T(y_1)  T(y_2)] \> &=&  \frac{1}{2} \sum_{\mathrm{perms} \ \{a_i\}} \Big( \< T(z_{a_1}) T(z_{a_2}) T(y_1) \> \<T(z_{a_3})  T(y_2) \> 
\nn \\
 &+&  \< T(z_{a_1}) T(z_{a_2}) T(y_2) \> \<T(z_{a_3})  T(y_1) \> \Big).
\ee
Our prescription leads to correlators that do not have any singularities as $y_i \to y_j$, so it certainly regulates divergences.  
In the next two sections we provide evidence that the prescription agrees with the Virasoro Ward identity.

\subsubsection{Test with  Two Stress Tensors}
\label{app:Testwith2Ts}

The version of the Ward identity with one stress tensor, namely 
\be
\left\< T(z_1) W_h(z;0) \right\> = \< T(z_1) \CO_h(z) \CO_h(0) \>,
\ee
can be easily verified.  According our regulator it only receives contributions from a single stress tensor from $W_h$.  Thus let us proceed to study the two stress tensor case.  For convenience we define
\be
F \equiv \left\< T(z_1) T(z_2) W_h(z;0) \right\> \overset{?}{=} \< T(z_1) T(z_2) \CO(z) \CO(0) \> .
\ee
Let us show explicitly that the second equality holds.  This means we must show that
\be
z^{2h} F &=& \frac{c}{2 (z_1 - z_2)^4  } 
\\ 
&& + \frac{h^2 z^{4}}{\left(z-z_1\right){}^2 z_1^2
   \left(z-z_2\right){}^2 z_2^2} + \frac{2 h z^{2 } }{\left(z-z_1\right) z_1 \left(z-z_2\right)
   \left(z_1-z_2\right){}^2 z_2} .
\nn 
\ee
The contribution on the first line just arises trivially from the leading, $T$-independent, term in the expansion of $W$ in powers of $T$ since it is just $\<TT\> \<W\>$, so let us move on to the other terms.  

We will compute the second line from the regulated Wilson line formalism.  
We receive contributions from both $\<T(z_1) T(z_2) [T(y_1) ]\>$ and $\<T(z_1) T(z_2) [T(y_1) T(y_2)]\>$ correlators.  In fact the order $h^2$ piece of $F_2$ only receives contributions from the latter, and the correct result follows easily due to the form of the integration kernel from equation (\ref{eq:vacblockf2kernel}).  This accords with the fact that the order $h^2$ piece is really just a product of $\<T \CO \CO\>$ correlators divided by a normalization $\<\CO \CO\>$.

Thus let us focus on the piece of $F$ proportional to $h$, which gets contributions from both types of regulated correlator.  The first contribution is
\be
z^{2h} F_{[T],h} & = &  \frac{1}{c} \int_0^z y \frac{12 h}{z} (z-y)(y) \left\< T(z_1) T(z_2) [T(y)] \right\>
\nn \\
&=& \frac{12 h}{z}   \int_0^z dy  \frac{ (z-y)(y)}{(z_1 - y)^2 (z_2-y)^2 (z_1 - z_2)^2}.
\ee
The integral is non-trivial and produces both rational functions and logarithms.  The other contribution is
\be
z^{2h} F_{[TT],h} =  \frac{36 h}{z^2 c^2} \int_0^z d y_1 d y_2 (z - \max(y_1,y_2))^2 (\min(y_1,y_2))^2 \left\< T(z_1) T(z_2)  [T(y_1) T(y_2) ] \right\>.
\nn
\ee
The regulated correlator only includes disconnected pieces.  Once again the integrals are rather non-trivial, and produce both rational functions and logarithms.  However, the two pieces sum to provide the correct result
\be
z^{2h} \left( F_{[T], h} + F_{[TT],h} \right) = \frac{2 h z^2}{\left(z-z_1\right) z_1 \left(z-z_2\right)
   \left(z_1-z_2\right){}^2 z_2},
\ee
as desired.  By the definition of the regulator in section \ref{app:TheRegulator}, there are no other contributions.  Note that the free boson regulator would also produce extra terms that contaminate the Ward identity at higher orders in $1/c$.

\subsubsection{Test of the Regulator to Higher Orders}
\label{app:OrderbyOrderTest}
 
In this section we will explain how we have tested our regulator to higher orders.  The basic idea is to compare the prescription for the Wilson line against the OPE block order-by-order in a small $z$ expansion, as in eq. (\ref{eq:OPEseries}).  We emphasize that this is a check of the Wilson line as an operator, since the OPE block can be inserted inside correlators with arbitrary additional local operators.  To organize the Wilson line in a small $z$ expansion, it is useful to use the expression (\ref{eq:PathIntegralWilsonLine}), reproduced here for convenience:
\begin{equation}
\< h| W( z_f, z_i)| h\> =
\left( e^{\int_{z_i}^{z_f} dz \frac{12 T(z)}{c} x(z)}\frac{1}{x(z_i)^2} \right)^h , 
 \quad -x'(z) = 1 + \frac{6 T(z)}{c} x^2(z), \quad x(z_f) =0.
\end{equation}
It is straightforward to solve for $x(z)$ in an expansion around $z \sim z_f$:
\be
x(z) = (z_f-z) + \frac{2 (z_f-z)^3 T(z_f)}{c} - \frac{3(z_f-z)^4 T'(z_f)}{c} + \dots .
\ee
Substituting back into the expression for $W_h(z_f, z_i)$, we find an expansion for $W_h$ at small $z$:
\begin{equation}
z^{2h} W_h(0, z) = 1 + \frac{2h T(0) }{c} z^2 + \frac{h T'(0)}{c} z^3 + \frac{ h(4(1+5h)[T^2(0)]+ 3 c T''(0))}{10 c^2} z^4 + \CO(z^5), 
\label{eq:WLseries} 
\end{equation}
where we have taken $z_f \rightarrow 0, z_i \rightarrow z$. This expansion has the advantage that at each order in $z$, we find a finite number of new multi-$T$ operators, and the weight of those operators is fixed by the power of $z$ where they appear.  So for instance, at $\CO(z^2)$, the only operator is $T(0)$, whereas at $\CO(z^4)$, there are two operators, $T^2(0)$ and $T''(0)$, both with weight 4.  The regulated product $[T^2(0)]$ is defined in accordance with our proposal (\ref{eq:RegProp}). To compare (\ref{eq:WLseries}) with the OPE block (\ref{eq:OPEseries}) it is supposed to reproduce, we have to convert the stress tensor and its regulated products into Virasoro modes. For a single stress tensor $T(0)$ and its derivatives $T^{(n)}(0)$, the conversion is the standard one:
\be
T^{(n)}(0) = n! L_{-(n+2)}.
\ee
For the regulated products, the conversion involves some work: we have to turn our prescription for the correlators of $[T^2(0)]$ with products of $T(z)$ into an expression for $[T^2(0)]$ itself.  Since $[T^2(0)]$ has weight 4, it must be a linear combination of $L_{-2}^2$ and $L_{-4}$, and therefore it is sufficient to look at its overlap with $T(z)$ and $T(z_1)T(z_2)$.  According to our proposal,
\be
\< T(z) [T^2(y)]\> &=& 0, \nn\\
\< T(z_1) T(z_2) [T^2(y)]\> &=& \frac{c^2}{2 (z_1-y)^4 (z_2-y)^4}. 
\ee
These conditions uniquely fix the coefficients of $L_{-2}^2$ and $L_{-4}$:
\be
[T^2(0)] &=& \frac{c (5 L_{-2}^2 - 3 L_{-4})}{22+5c}.
\label{eq:RegTsq}
\ee
Now, by direct substitution into (\ref{eq:WLseries}), one can compare with the OPE block (\ref{eq:OPEseries}) and see by inspection that they agree up to $z^4$. In fact it is not hard to see the converse is also true; demanding that (\ref{eq:WLseries}) and (\ref{eq:OPEseries}) agree at $\CO(z^4)$ uniquely fixes the regulated product $[T^2(0)]$ to be (\ref{eq:RegTsq}).  Since $[T^2(0)]$ acts at the origin, the above expression should be understood to always act to the right of any other operator insertions; however $[T^2(0)]$ can be translated in a straightforward manner to $[T^2(y)]$ at any other point using global conformal generators.\footnote{For instance, in terms of the conventional operator $(TT)(z)  \equiv \oint \frac{dw}{2\pi i} \frac{T(w) T(z)}{w-z}$ made from contour integration,
$[T^2(y)]$ is just $[T^2(y)] = \left( 1+ \frac{22}{5c} \right)^{-1} \left( (TT)(y) - \frac{3}{10} T''(y) \right)$. }

Clearly, we can continue this prescription to arbitrarily high orders in $z$, checking at each order that the OPE block is correctly reproduced.  At order  $z^n$, regulated products of weight $n$ appear, and so it is sufficient to inspect their overlap with powers of $T$ and their derivatives up to weight $n$; or, in other words, it is sufficient to inspect their overlap with Virasoro descendants of the vacuum at level $n$. At $\CO(z^5)$, the only regulated operator that appears is $[T(0)T'(0)]$, which is a global descendant of $[T^2(0)]$.  So while it is important (and true) that our proposal correctly reproduces the OPE block at $\CO(z^5)$, this was guaranteed by the agreement at level 4 if we assume that global descendants are treated correctly by the Wilson line prescription.

   At level 6, our proposal for $\< T(z_1) \dots T(z_n) [T(y_1) \dots T(y_m)]\>$  can be summarized, after a short calculation, by
\be
 \< L_2^3 [T^3(0)]\> = \frac{3c^3}{4} , \qquad 
\< L_3^2 [T'^2(0)]\> =8c^2  .
\ee
All other overlaps between $[T^3(0)]$ and $[T'^2(0)]$  and level 6 vacuum descendants vanish according to the proposal (\ref{eq:RegProp}). 
As we did with $[T^2(0)]$, one can convert these conditions into expressions for $[T^3(0)]$ and $[T'^2(0)]$:
\be
\left[ T'^2(0)\right] &=& \frac{992 c L_{-2}^3 + 2c (512+5c(80+7c))L_{-3}^2 - 248c (16+c)L_{-4}L_{-2} -8c(160+c(94+9c))L_{-6}}{(-1+2c)(22+5c)(68+7c)} ,
\nn\\
\left[T^3(0)\right] &=&  \frac{c^2 (29+70c)L_{-2}^3 + 93 c^2 L_{-3}^2 - 3 c^2 (67+42 c)L_{-4}L_{-2} - 6 c^2 (13+10c) L_{-6}}{(-1+2c)(22+5c)(68+7c)} .\nn\\
\ee
Other regulated operators that appear at level 6, such as $[T''(0) T(0) + T'^2(0)]$, are global descendants of lower level operators. Substituting the resulting expressions for the regulated operators at level 6 into the Wilson line (\ref{eq:WLseries}) at $\CO(z^6)$, one correctly reproduces the OPE block at $\CO(z^6)$.  

We do not have a proof that this agreement continues to all orders.  However, we have checked the agreement explicitly up to level 10.  The additional independent overlaps that are needed up to this order are predicted by our proposal (\ref{eq:RegProp}) to be
\be
\textrm{level 8} &:& \nn\\
&& \< L_2^4 [ T^4 ] \> = \frac{3 c^4}{2} , \qquad \< L_2 L_3^2 [ T T'^2]\> = 4 c^3, \nn\\
&& \< L_2^4 [T''^2]\> = 216 c^2, \qquad \< L_2^2 L_4 [T''^2]\> = 120 c^2 , \qquad \< L_4^2 [T''^2]\> = 200 c^2 , \nn\\
\textrm{level 9} &:&\nn\\
 &&  \< L_3^3 [ T'^3]\> = 48 c^3 , \nn\\
\textrm{level 10} &:& \nn\\
 && \< L_2^5 [T^5]\> = \frac{15 c^5}{4} , \qquad \< L_2^2 L_3^2 [T^2 T'^2]\> = 4 c^4, \nn\\
&&\< L_2^5 [T T''^2]\> = 540 c^3, \qquad \< L_2^3 L_4 [T T''^2]\> = 180 c^3 , \qquad \< L_2 L_4^2 [T T''^2]\> = 100 c^3, \nn\\
&&\< L_2^2 L_3^2 [T'''^2]\> = 2304 c^2 , \qquad \< L_2 L_3 L_5 [T'''^2]\> = 2880 c^2, \qquad \< L_5^2 [T'''^2]\> = 7200 c^2. \nn\\
\ee 
All $T$s here are implicitly at $y=0$. A direct, brute force computation shows that these predict the Wilson line in agreement with the OPE block up to $\CO(z^{10})$.  

As a final comment, we emphasize that while the correlators of the form \\ $\< T(z_1) \dots T(z_n) [ T(y_1) \dots T(y_m)] \>$ are fairly simple, the correlators of multiple regulated operators are not. This is because the procedure of translating the former into operator equations for the regulated products introduces complicated expressions; one can think of this as inserting a projector that brings the regulated products back into the space of Virasoro descendants of the vacuum.  As an example, one can compute the following correlator of $[T'^2]$ with itself:
\be
\< [T'(\infty) T'(\infty)][T'(0) T'(0)]\> &=& \frac{16 c^3 \left(35 c^2+400 c+512\right)}{(2 c-1) (5 c+22) (7 c+68)}.
\ee
An unfortunate consequence is that  a direct calculation of the Virasoro blocks at higher orders in $1/c$ will be rather involved, likely more so than existing methods for computing Virasoro blocks.  Even at a conceptual level, it would be preferable to have a prescription that applies directly to correlators of multiple regulated operators, unlike the indirect prescription proposed here.

\bibliographystyle{utphys}
\bibliography{VirasoroBib}

\end{document}